\documentclass[structabstract]{aa}  
\usepackage{graphicx}
\usepackage{color}
\usepackage{epsfig}
\usepackage{supertabular}
\usepackage{longtable}
\usepackage{subfigure}
\usepackage{txfonts}
\usepackage{natbib}
\bibpunct{(}{)}{;}{a}{}{,}
\newcommand{\kms}{km~s$^{-1}$}
\usepackage{booktabs}
\begin{document}
   \title{The early-type dwarf galaxy population of the Centaurus cluster\thanks{Based on observations obtained at the European Southern Observatory, Chile (Observing Programme 67.A-0358).}}


   \author{I. Misgeld\inst{1,2} \and M. Hilker\inst{1} \and S. Mieske\inst{3}}


   \institute{European Southern Observatory, Karl-Schwarzschild-Strasse 2, 85748 Garching bei M\"unchen, Germany \\
	\email{imisgeld;mhilker@eso.org} \and Argelander Institut f\"ur Astronomie, Universit\"at Bonn, Auf dem H\"ugel 71, 53121 Bonn, Germany \and European Southern Observatory, Alonso de Cordova 3107, Vitacura, Santiago, Chile \\ \email{smieske@eso.org}
              }

   \date{}

 
  \abstract
   {} 
   {We present a photometric study of the early-type dwarf galaxy population of the Centaurus cluster, aiming at investigating the galaxy luminosity function (LF) and galaxy scaling relations down to the regime of galaxies with $M_V\sim -10$~mag.}
   {On deep VLT/FORS1 $V$- and $I$-band images of the central part of the cluster, we identify cluster dwarf-galaxy candidates using both morphological and surface brightness selection criteria. Photometric and structural parameters of the candidates are derived from analysis of their surface brightness profiles. Fundamental scaling relations, such as the colour--magnitude and the magnitude-surface brightness relation, are used to distinguish the cluster from the background.}
   {We find a flat LF with a slope of $\alpha = -1.14\pm0.12$ for $M_V>-14$ mag, when fitting a power law to the completeness-corrected galaxy number counts. Applying a Schechter function leads to a consistent result of $\alpha\sim-1.1$. When plotting the central surface brightness of a S\'ersic model vs. the galaxy magnitude, we find a continuous relation for magnitudes $-20<M_V<-10$ mag, with only the brightest core galaxies deviating from this relation, in agreement with previous studies of other clusters. Within our Centaurus dwarf galaxy sample we identify three very compact objects. We discuss whether they belong to the class of the so-called compact elliptical galaxies (cEs). In a size--luminosity diagram ($R_{\mathrm{eff}}$ vs. $M_V$) of early-type galaxies from a range of environments, we observe that $R_{\mathrm{eff}}$ slowly decreases with decreasing luminosity for $-21<M_V<-13$~mag and decreases more rapidly at fainter magnitudes. This trend continues to the ultra-faint Local Group dwarf galaxies ($M_V\sim -4$~mag).}
   {The continuous central surface brightness vs. absolute magnitude relation and the smooth relation in the size--luminosity diagram over a wide range of magnitudes are consistent with the interpretation of dwarf galaxies and more massive elliptical galaxies being one family of objects with gradually changing structural properties. The most massive core galaxies and the rare cE galaxies are the only exceptions.}


   \maketitle 
%

\section{Introduction}
Probing the faint end of the luminosity function (LF) in galaxy clusters and groups has in many cases exposed a discrepancy between the number of observed dwarf galaxies and the number of dark matter (DM) sub-haloes predicted by current hierarchical cold dark matter models -- the so-called missing satellites problem \citep{1999ApJ...522...82K}. Its origin is still a matter of debate. Either there are many faint satellites not yet discovered, the predictions of the hierarchical models are not reliable, or the large majority of low-mass DM haloes have not formed any stars. To quantify this discrepancy, the LF can be parametrised by the Schechter function, whose logarithmic faint-end slope $\alpha$ can be contrasted with the predicted slope of about $-2$ for the mass spectrum of cosmological DM haloes \citep[e.g.][]{1999ApJ...524L..19M, 2001MNRAS.321..372J}. The observed value of $\alpha$ is generally much lower than expected. This has been shown in many studies, for both low density environments like the Local Group (LG) and galaxy clusters \citep[e.g.][]{1999AJ....118..883P, 2002MNRAS.335..712T, 2005MNRAS.357..783T}. 

For the LG and the galaxy clusters Fornax, Perseus and Virgo, the faint-end slope of the LF can be determined by direct cluster membership assignment via spectroscopic redshift measurements \citep[e.g.][]{1999A&AS..134...75H, 2001ApJ...548L.139D,2008MNRAS.383..247P, 2008AJ....135.1837R}. For other galaxy clusters, however, only photometric data are available at magnitudes where $\alpha$ dominates the shape of the LF ($M_V\gtrsim -14$~mag). In this case, cluster galaxies have to be separated from background galaxies either by means of statistical background subtraction or by their morphology and correlations between global photometric and structural parameters. For the latter case, the colour--magnitude relation (CMR) can be used, which is observed not only for giant elliptical galaxies \citep[e.g.][]{1977ApJ...216..214V, 1997A&A...320...41K, 2006MNRAS.370.1106G}, but also for early-type \textit{dwarf} galaxies \citep[e.g.][]{1997PASP..109.1377S, 2003A&A...397L...9H, 2006A&A...459..679A, 2007A&A...463..503M, 2008AJ....135..380L, 2008A&A...486..697M}. 

Although they form a common relation in a colour--magnitude diagram, the question of whether giant elliptical galaxies, on the one hand, and early-type dwarf galaxies (dwarf ellipticals (dEs) and dwarf spheroidals (dSphs)), on the other, have the same origin, has been a controversial issue over the past decades. Two major perceptions exist. First, dwarf elliptical galaxies are not the low luminosity counterparts of giant elliptical galaxies, but rather an independent class of objects. This point of view is mainly based on studies of relations between galaxy surface brightness and magnitude, or surface brightness and size -- the Kormendy relation \citep[e.g.][]{1977ApJ...218..333K, 1985ApJ...295...73K, 1991A&A...252...27B, 1992ApJ...399..462B, 1993ApJ...411..153B}. These studies showed an apparent dichotomy of dwarf and giant elliptical galaxies, in the sense that for dwarfs the surface brightness increases with luminosity, whereas the opposite trend is seen for giants. Moreover, a weaker dependence of size on luminosity was observed for dwarfs than for giants. Recently, \citet{2008arXiv0810.1681K} reaffirmed those results in their study of a large sample of Virgo cluster early-type galaxies. They concluded that dwarf galaxies are structurally distinct from giant early-type galaxies and that different mechanisms are responsible for their formation (see also \citealt{2008A&A...489.1015B} and \citealt{2008ApJ...689L..25J}).

The alternative point of view is that the apparent dichotomy is the result of a gradual variation of the galaxy light profile shape with luminosity. If the light profile is described by the \citet{1968adga.book.....S} law, the different behaviour of dwarf and giant early-type galaxies in the surface brightness vs. magnitude relation compared to the Kormendy relation is a natural consequence of the linear relation between S\'ersic index $n$ and galaxy magnitude \citep[e.g.][]{1997ASPC..116..239J, 2003AJ....125.2936G, 2005A&A...430..411G, 2006ApJS..164..334F, 2007ApJ...671.1456C, 2008IAUS..246..377C}. This implies that dwarf ellipticals represent the low luminosity extension of massive elliptical galaxies.

Continuing a series of similar investigations in Fornax and Hydra\,I \citep{2003A&A...397L...9H, 2007A&A...463..503M, 2008A&A...486..697M}, we study in this paper the early-type dwarf galaxy population of the Centaurus cluster, aiming at the investigation of the galaxy LF and photometric scaling relations. It is based on deep VLT/FORS1 imaging data of the central part of the cluster. In Sect.~\ref{sec:sample} we describe the observations, the sample selection and the photometric analysis of the dwarf galaxy candidates. We present our results in Sect.~\ref{sec:results}. The findings are summarised and discussed in Sect.~\ref{sec:discussion}.

\subsection{The Centaurus cluster (Abell 3526)}
The Centaurus cluster, which is categorised as a cluster of richness class zero and Bautz-Morgen type I/II \citep{1958ApJS....3..211A, 1970ApJ...162L.149B}, is the dominant part of the Centaurus-Hydra supercluster \citep{1986AJ.....91....6D, 1987AJ.....93.1338D}. It is characterised by an ongoing merger with an in-falling sub-group and irregular X-ray isophotes \citep{1999ApJ...520..105C, 2001PASJ...53..421F}. The main cluster component is Cen30 with NGC 4696 at its dynamical centre, whereas Cen45 is the in-falling sub-group with NGC 4709 at its centre \citep{1986MNRAS.221..453L, 1997A&A...327..952S}.

\citet{2005A&A...438..103M} derived the distance to Centaurus by means of surface brightness fluctuation (SBF) measurements. They found the SBF-distance to be $45.3\pm2.0$~Mpc ($(m-M)=33.28\pm 0.09$~mag). \citet{2001ApJ...546..681T}, however, measured a significantly shorter distance of 33.8~Mpc, which may partially be attributed to selection effects \citep[see discussion in][]{2003A&A...410..445M}. For 78 cluster galaxies, a mean redshift of $v_{\mathrm{rad}}=3656$~\kms~is determined in \citet{2006AJ....132..347C}. This corresponds to a distance of $50.8\pm 5.6$~Mpc, assuming $H_0 = 72\pm 8$~\kms~Mpc$^{-1}$ \citep{2001ApJ...553...47F}, and agrees with the SBF-distance within the errors. Throughout this paper, we adopt a distance modulus of $(m-M)=33.28$~mag \citep{2005A&A...438..103M}, which corresponds to a scale of 220 pc/arcsec.

\section{Observations and sample selection}
\label{sec:sample}
\begin{figure}
	\resizebox{\hsize}{!}{\includegraphics{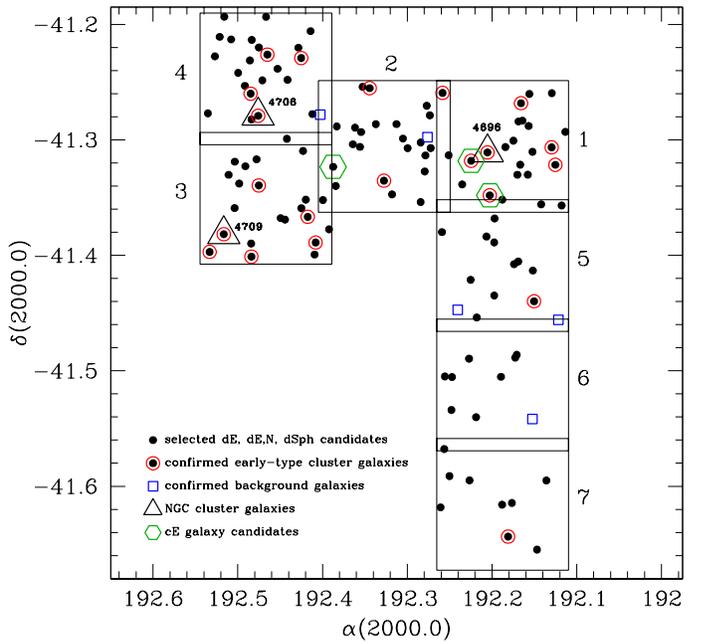}}
	\caption{Map of the seven VLT/FORS1 cluster fields (large open squares) with the selected dwarf galaxy candidates, the spectroscopically confirmed cluster members and background galaxies. The major cluster galaxies NGC 4696, NGC 4709 and NGC 4706 are marked by open triangles. Green open hexagons are compact elliptical galaxy (cE) candidates (see Sect.~\ref{sec:cEs}).}
	\label{fig:fields}
\end{figure}

The observations were executed in a service mode run at the Very Large Telescope (VLT) of the European Southern Observatory (ESO, programme 67.A-0358). Seven fields in the central part of the Centaurus cluster of size  $7'\times7'$ were observed in Johnson $V$ and $I$ filters, using the instrument FORS1 in imaging mode. The fields cover the central part of the Centaurus cluster with its sub-components Cen30 and Cen45, which are centred on NGC 4696 and NGC 4709, respectively (see Fig.~\ref{fig:fields}). The exposure time was $4\times373$ s in $V$ and $9\times325$ s in $I$. The seeing was excellent, ranging between $0.4''$ and $0.6''$. Additional short images (30 s in both filters) were taken to be able to analyse the brightest cluster galaxies, which are saturated on the long exposures. Furthermore, an eighth (background) field located about $2.5\degr$ west of NGC 4696 was observed.

\begin{figure*}
	\centering
	\subfigure{\includegraphics[width=4.5cm]{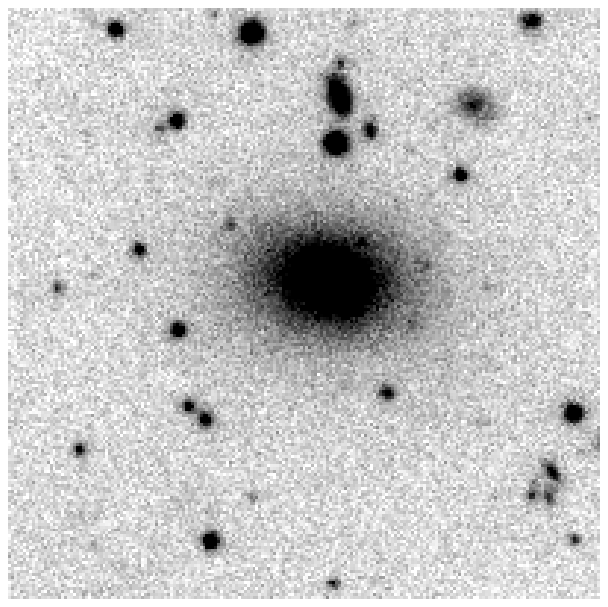}}
	\subfigure{\includegraphics[width=4.5cm]{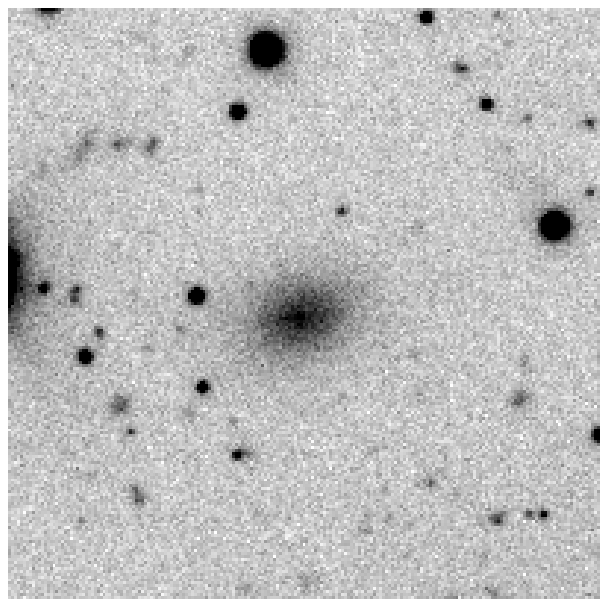}}
	\subfigure{\includegraphics[width=4.5cm]{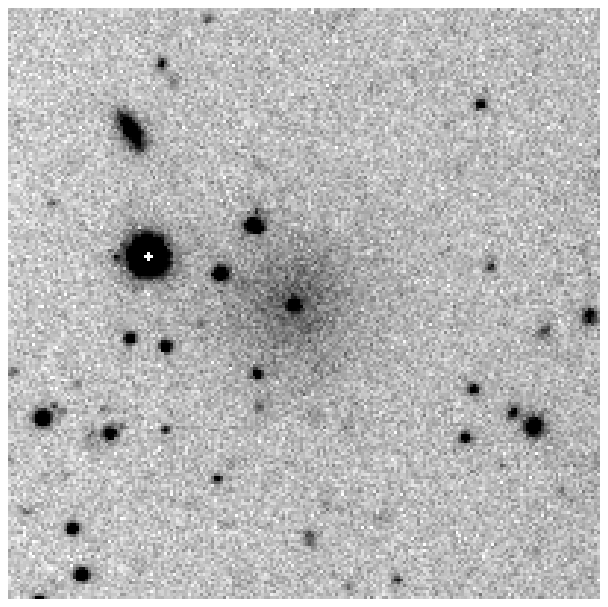}}
	\subfigure{\includegraphics[width=4.5cm]{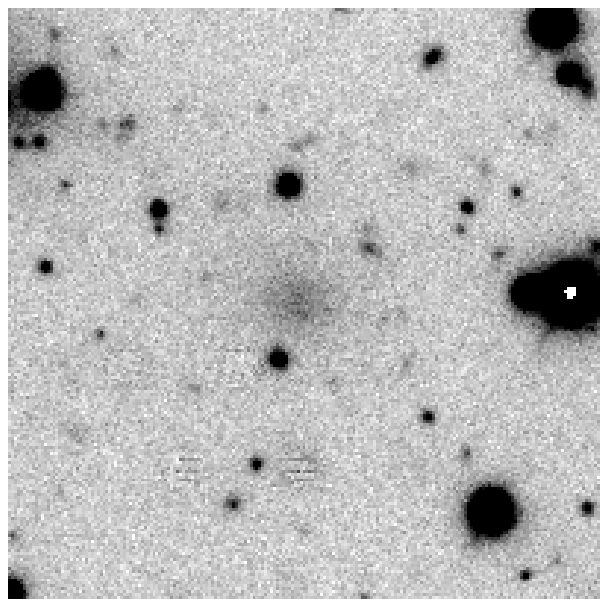}}
	\caption{Thumbnail images of four cluster dwarf galaxy candidates that fulfil our selection criteria (two dEs, one dE,N and one dSph). The objects' absolute magnitudes from the left to the right are: $M_V=-14.4, -13.3, -13.2, -12.0$ mag, assuming a distance modulus of $(m-M)=33.28$ mag \citep{2005A&A...438..103M}. The thumbnail sizes are $40''\times40''$ ($8.8\times 8.8$ kpc at the cluster distance).}
	\label{fig:candidates}
\end{figure*}

In this study, we are interested in early-type galaxies. They were selected based on morphology and spectroscopic redshifts. Our images contain 21 spectroscopically confirmed early-type cluster galaxies and one late-type (Sc) cluster galaxy \citep{1997A&AS..124....1J, 1997A&A...327..952S, 2007ApJS..170...95C, 2007A&A...472..111M}. The cluster membership criterion was adopted to be $1700<v_{\mathrm{rad}}<5500$~\kms.

In order to identify new early-type dwarf galaxy candidates on the images, we followed the same strategy as in our investigations of the dwarf galaxy populations in Fornax and Hydra\,I \citep{2003A&A...397L...9H, 2007A&A...463..503M, 2008A&A...486..697M}. It is a combination of visual inspection and the use of SExtractor \citep{1996A&AS..117..393B} detection routines. We first added several simulated Local Group (LG) dEs and dSphs (projected to the Centaurus cluster distance) to the images. Their magnitudes and central surface brightnesses were adopted according to the relations found by \citet{2003AJ....125.1926G} and \citet{2006MNRAS.365.1263M}. Afterwards, the images were inspected by eye, and candidate cluster dwarf galaxies were selected by means of their morphological resemblance to the simulated galaxies. The main selection criterion was a smooth surface brightness distribution and the lack of substructure or spiral arms. This first search resulted in the identification of 89 previously uncatalogued dE/dSph candidates, from which four are shown in Fig.~\ref{fig:candidates}.

In a second step, we used the SExtractor detection routines to quantify the detection completeness in our data (see also Sect. \ref{sec:lumfunction}) and to find more dwarf galaxy candidates, in particular at the faint magnitude and surface brightness limits. The detection sensitive SExtractor parameters were optimised such that a maximum number of objects from the by-eye catalogue was recovered by the programme. Only 12 of the 89 obvious by-eye detections were not recovered, mostly due to their position close to another bright object or close to the image border. In our search for new dwarf galaxy candidates we focused on those sources in the SExtractor output catalogue whose photometric parameters matched the parameter range of the simulated dwarf galaxies. For this, we applied cuts in the SExtractor output-parameters \texttt{mupeak}, \texttt{area} and \texttt{fwhm} to constrain the output parameter space to the one found for the simulated LG dwarf galaxies \citep[see also][]{2008A&A...486..697M}. We thus rejected barely resolved and apparently small objects with high central surface brightnesses, both being likely background galaxies. The applied cuts are described in detail in Sect. \ref{sec:lumfunction}. In this way, 8 additional objects in the magnitude range $-11.0<M_V<-9.4$ mag were found and added to the by-eye catalogue. On the background field, neither the visual inspection nor the SExtractor analysis resulted in the selection of an object.

To our photometric sample we also added five spectroscopically confirmed background early-type galaxies that are located in the observed fields, in order to be able to compare their photometric properties with the ones of the objects in the by-eye catalogue. In total, our sample contains 123 objects, for which Fig.~\ref{fig:fields} shows a coordinate map.

\subsection{Photometric analysis}
For each selected object we created thumbnail images with sizes extending well into the sky region (see Fig. \ref{fig:candidates}). On these thumbnails we performed the sky subtraction and fitted elliptical isophotes to the galaxy images, using the IRAF-task \texttt{ellipse} in the \texttt{stsdas}\footnote{Space Telescope Science Data Analysis System, STSDAS is a product of the Space Telescope Science Institute, which is operated by AURA for NASA.} package. During the fitting procedure the centre coordinates, the position angle and the ellipticity were fixed, except for some of the brightest cluster galaxies ($V_0\lesssim 15.5$~mag) where the ellipticity or both the ellipticity and the position angle considerably changed from the inner to the outer isophotes. In those cases one or both parameters were allowed to vary.

The total apparent magnitude of each object was derived from a curve of growth analysis. The central surface brightness was determined by fitting an exponential as well as a \citet{1968adga.book.....S} law to the surface brightness profile. From the fit we excluded the inner $1''$ (about 1.5 seeing disks) and the outermost part of the profile, where the measured surface brightness was below the estimated error of the sky background. Corrections for interstellar absorption and reddening were taken from \citet{1998ApJ...500..525S}, who give $A_V=0.378$~mag and $E(V-I)=0.157$~mag for the coordinates of NGC 4696. We adopt these values for all of our observed fields. Zero points, extinction coefficients and colour terms for the individual fields and filters are listed in Table \ref{tab:photcal} in the appendix (only available on-line).

\section{Global photometric and structural parameters}
\label{sec:results}
In this section the results of the photometric analysis are presented. In Sect.~\ref{sec:scalings} we address the colour--magnitude and the magnitude-surface brightness relation of the Centaurus early-type dwarf galaxies and use these relations to facilitate the distinction of cluster and background galaxies. The galaxy luminosity function of probable cluster members is studied in Sect.~\ref{sec:lumfunction}. The structural parameters of the cluster galaxies, as obtained from S\'ersic fits to the surface brightness profiles, are presented in Sect.~\ref{sec:sersic}. Table~\ref{tab:Centaurussample} in the appendix summarises the obtained photometric parameters of the 92 probable Centaurus cluster early-type galaxies in our sample (only available on-line).

\subsection{Fundamental scaling relations}
\label{sec:scalings}
\begin{figure}
	\resizebox{\hsize}{!}{\includegraphics{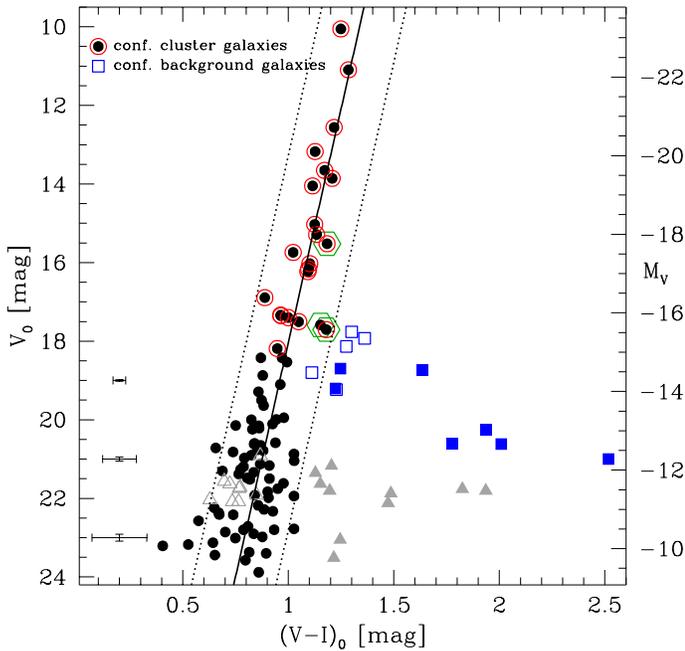}}
	\caption{Colour--magnitude diagram of early-type galaxies in the Centaurus cluster. Black dots are probable cluster galaxies, selected by their morphology. Red open circles (blue open squares) mark spectroscopically confirmed cluster members (background galaxies). Blue filled squares and grey filled triangles are probable background objects (see text for details). Grey open triangles are objects which do not follow the magnitude-surface brightness relation (cf. Fig.~\ref{fig:magmu}). Green open hexagons mark the candidates for compact elliptical galaxies (see Sect.~\ref{sec:cEs}). Typical errorbars are indicated on the left. The solid line is a linear fit to the cluster member candidates (Eq.~(\ref{eq:cmr})) with its $2\sigma$ deviations (dotted lines).}
	\label{fig:cmd}
\end{figure}

Figure \ref{fig:cmd} shows a colour--magnitude diagram of our sample of early-type galaxies, as defined in Sect. \ref{sec:sample}. Spectroscopically confirmed cluster galaxies ($V_0\lesssim18$ mag) form a colour--magnitude relation (CMR) in the sense that brighter galaxies are on average redder. This sequence continues down to the faint magnitude limit of our survey ($M_V\sim -10$ mag), which is comparable to the absolute magnitudes of the LG dwarf galaxies Sculptor and Andromeda III \citep{2003AJ....125.1926G, 2006MNRAS.365.1263M}. The larger scatter at faint magnitudes is consistent with the larger errors in $(V-I)_0$. The mean measured error in $(V-I)_0$ is 0.03, 0.08 and 0.13~mag for the three magnitude intervals indicated in Fig.~\ref{fig:cmd}. The intrinsic scatter of the datapoints in the same intervals is 0.06, 0.09 and 0.14~mag, respectively, only marginally larger than the measurement errors. Our data do therefore not require an increase of metallicity or age spread among individual galaxies at faint luminosities, compared to brighter luminosities. For the linear fit, we weighted each data point by its colour error, resulting in:
\begin{equation}
\label{eq:cmr}
 (V-I)_0 = -0.042(\pm0.001) \cdot M_V + 0.33(\pm0.02)
\end{equation}
with a rms of 0.10. This is in good agreement with the CMRs observed in Fornax and Hydra\,I \citep{2007A&A...463..503M, 2008A&A...486..697M}. Table~\ref{tab:relations} lists the CMR coefficients for each of those clusters. For consistency, we re-fitted the Hydra\,I data using the error weighted values, which slightly changes the coefficients given in \citet{2008A&A...486..697M}.

\begin{figure}
	\resizebox{\hsize}{!}{\includegraphics{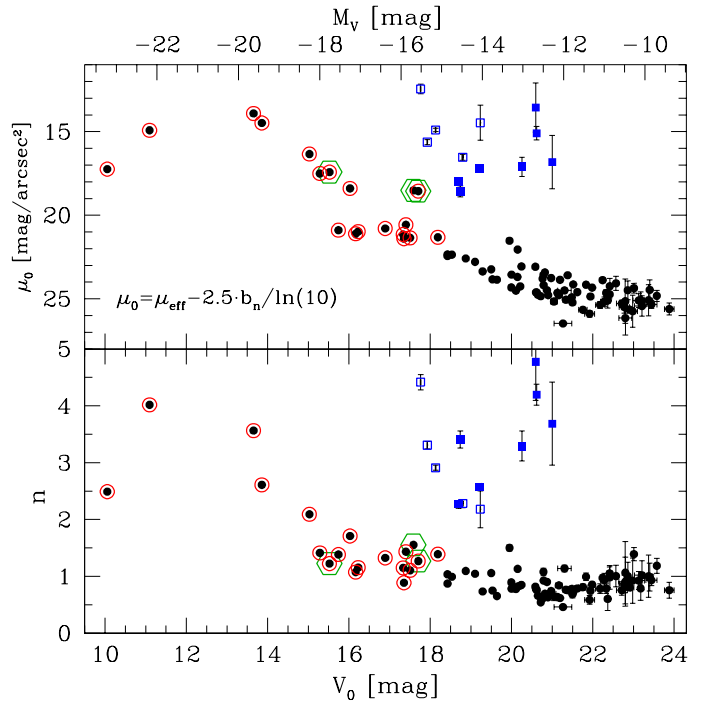}}
	\caption{Parameters of the S\'ersic fits to the galaxy surface brightness profiles. The top (bottom) panel shows the central surface brightness $\mu_0$ (profile shape index $n$) plotted vs. the galaxy magnitude. Black dots are all galaxies that were considered cluster members. Spectroscopically confirmed cluster galaxies are marked by red open circles. Blue open (filled) squares are confirmed (likely) background galaxies (cf. Fig.~\ref{fig:cmd}). The green open hexagons mark the three compact elliptical galaxy candidates (see Sect.~\ref{sec:cEs}).}
	\label{fig:sersic}
\end{figure}

The CMR can be used as a tool to distinguish cluster from background galaxies. This is important, since the selection of cluster galaxy candidates solely based on morphological criteria can lead to the contamination of the sample with background objects that only resemble cluster dwarf ellipticals. In the bright magnitude range the cluster galaxies are identified by their redshift. In the intermediate magnitude range ($17.8<V_0<21.0$ mag), however, seven objects turn out to be likely background galaxies, although they passed our morphological selection criteria (filled squares in Fig.~\ref{fig:cmd}). All seven arguable objects have \citet{1948AnAp...11..247D} surface brightness profiles (also known as $R^{1/4}$ profiles), typical of giant elliptical galaxies. Five of those objects are too red to be a galaxy at $z\sim 0$, the other two share their position in the CMD with spectroscopically confirmed background galaxies (open squares in Fig.~\ref{fig:cmd}). Moreover, Fig.~\ref{fig:sersic} shows that the confirmed background galaxies as well as the seven likely background objects clearly differ from the cluster galaxies, because of their high central surface brightness and their large S\'ersic index. We consider 10 more morphologically selected objects with $V_0>21$ mag likely background objects, as their colours are significantly redder than those of other objects in the same magnitude range (see the filled triangles in Fig.~\ref{fig:cmd}).

\begin{figure}
	\resizebox{\hsize}{!}{\includegraphics{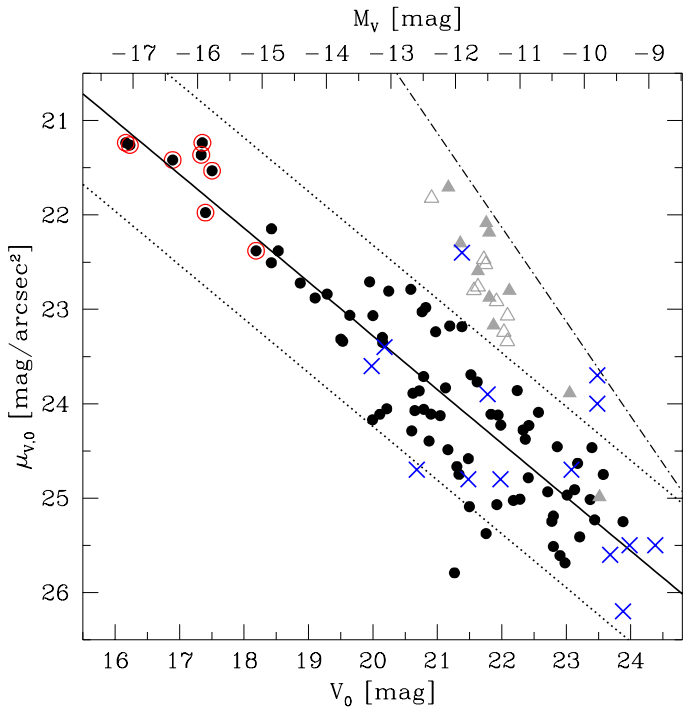}}
	\caption{Plot of the central surface brightness $\mu_{V,0}$, as derived from fitting an exponential law to the surface brightness profile, vs. the apparent magnitude $V_0$ for identified cluster dwarf galaxy candidates. Symbols are as in Fig.~\ref{fig:cmd}. Errors are comparable to the symbol sizes. The solid line is a linear fit to the black dots (Eq.~(\ref{eq:magmu})). Dotted lines are the $2\sigma$ deviations from the fit. Local Group dEs and dSphs, projected to the Centaurus distance, are given by the blue crosses (data from \citet{2003AJ....125.1926G} and \citet{2006MNRAS.365.1263M}). A scale length of $0.6''$ for an exponential profile, representing the resolution limit of our images, is indicated by the dash-dotted line.}
	\label{fig:magmu}
\end{figure}

We were not able to measure a colour for two objects in our sample (C-3-30 and C-1-47, see Table~\ref{tab:Centaurussample}), since they were located close to the image borders, only fully visible on the $V$-band images. However, they have a typical dE morphology and they fall onto the magnitude-surface brightness relation (Fig.~\ref{fig:magmu}). We thus treat them as probable cluster dwarf galaxies in the following analyses.

In Fig.~\ref{fig:magmu}, the central surface brightness $\mu_{V,0}$ is plotted against the apparent magnitude $V_0$ for all objects in our sample, whose surface brightness profiles are well represented by an exponential law. These are all objects with $V_0>16.1$ mag (see Table~\ref{tab:Centaurussample}), except for two compact elliptical galaxy candidates, whose properties we will discuss in Sect.~\ref{sec:cEs}. A linear fit to the probable cluster galaxies (black dots) leads to:
\begin{equation}
\label{eq:magmu}
 \mu_{V,0} = 0.57(\pm0.07) \cdot M_V + 30.90(\pm0.87)
\end{equation}
with a rms of 0.48. Given that the scatter in the data is much larger than the measured errors in both $M_V$ and $\mu_{V,0}$, we do not error weight the data points. The fit errors were derived from random re-sampling of the data points within their measured scatter. The same method was used in \citet{2007A&A...463..503M} for the Fornax dwarfs, and we re-analyse the Hydra\,I data \citep{2008A&A...486..697M} in the same way. The magnitude-surface brightness relations of the three clusters agree within the errors (see Table~\ref{tab:relations}).

When projected to the Centaurus distance, LG dwarf galaxies are mostly consistent with the same relation, with a few slightly more compact objects \citep{2003AJ....125.1926G, 2006MNRAS.365.1263M}. The likely background objects from Fig.~\ref{fig:cmd} (grey filled triangles) do not follow the relation, but they have central surface brightnesses about 1~mag/arcsec$^2$ higher than other objects of the same magnitude.

\begin{figure}
	\resizebox{\hsize}{!}{\includegraphics{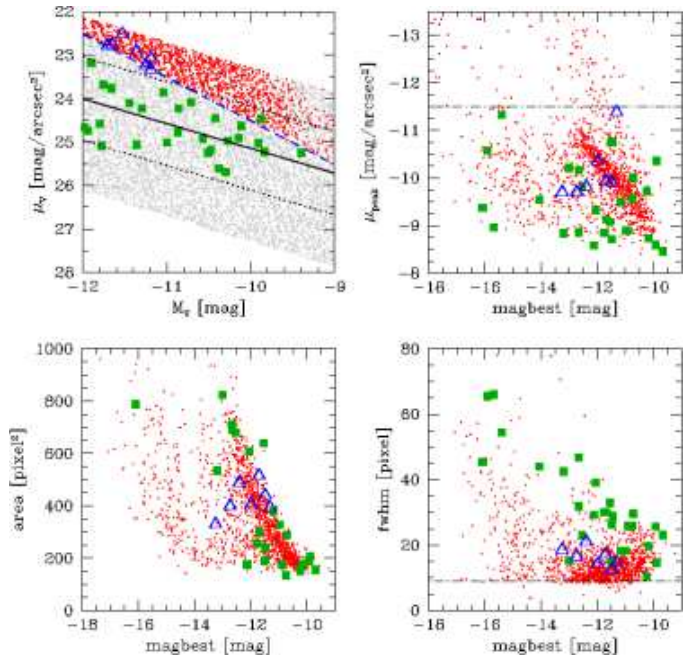}}
	\caption{SExtractor output-parameters of recovered simulated galaxies with an exponential scale length $<1''$ (red dots). \textit{The upper left panel} shows the input parameters absolute magnitude $M_V$ and central surface brightness $\mu_V$ of the artificial galaxies (grey dots), together with the probable cluster dwarf galaxies (green solid squares) that were recovered by SExtractor. Equation (\ref{eq:magmu}) with its $2\sigma$ deviations is plotted as in Fig.~\ref{fig:magmu}. The blue dashed line indicates a scale length of $1''$ for an exponential profile. Blue open triangles are the questionable objects discussed in Sect.~\ref{sec:scalings}. The SExtractor output-parameter \texttt{magbest} is plotted against \texttt{mupeak} (\textit{upper right}), \texttt{area} (\textit{lower left}) and \texttt{fwhm} (\textit{lower right}). Dash-dotted lines indicate the global cuts on \texttt{mupeak} and \texttt{fwhm} (see text for details).}
	\label{fig:artgal}
\end{figure}

An interesting sub-group of morphologically selected objects is marked by the open triangles. These nine objects are rather compact, having exponential scale lengths of $\lesssim 1''$, close to the resolution limit of our images. Although they lie on the cluster CMR (see Fig.~\ref{fig:cmd}), they are located more than $2\sigma$ away from the magnitude-surface brightness relation -- just as the likely background galaxies that were identified by their position aside the CMR. This suggests that these nine questionable objects are in fact background galaxies with colours similar to the cluster galaxies. However, the LG dwarf galaxy Leo\,I \citep{2003AJ....125.1926G} falls into the same parameter range (Fig.~\ref{fig:magmu}). If these objects were actual cluster galaxies, they would account for about 10\% of the whole dwarf galaxies population in our sample. Given their uncertain nature, we will analyse in Sect.~\ref{sec:lumfunction} how they affect the shape of the galaxy luminosity function. Ultimately, it remains to be clarified by spectroscopic measurements, whether they represent a family of rather compact early-type cluster members or background galaxies.

\subsection{The dwarf galaxy luminosity function}
\label{sec:lumfunction}
In order to study the faint-end of the galaxy luminosity function, the detection completeness in our data has to be quantified and the galaxy number counts have to be corrected to that effect. For this, 10\,000 simulated dwarf galaxies were randomly distributed in 500 runs in each of the seven cluster fields, using a C++ code. The background field was left out from this analysis, since we did not identify any potential dwarf galaxy. The upper left panel of Fig.~\ref{fig:artgal} illustrates the input-parameter range of the simulated galaxies, which extends well beyond the observed parameter space. The artificial galaxies were then recovered by SExtractor, and the SExtractor output-parameters \texttt{magbest}, \texttt{mupeak}, \texttt{area} and \texttt{fwhm} were compared with the parameters of the sample of probable cluster dwarf galaxies, as defined in Sect.~\ref{sec:scalings}.

\begin{figure}
	\resizebox{\hsize}{!}{\includegraphics{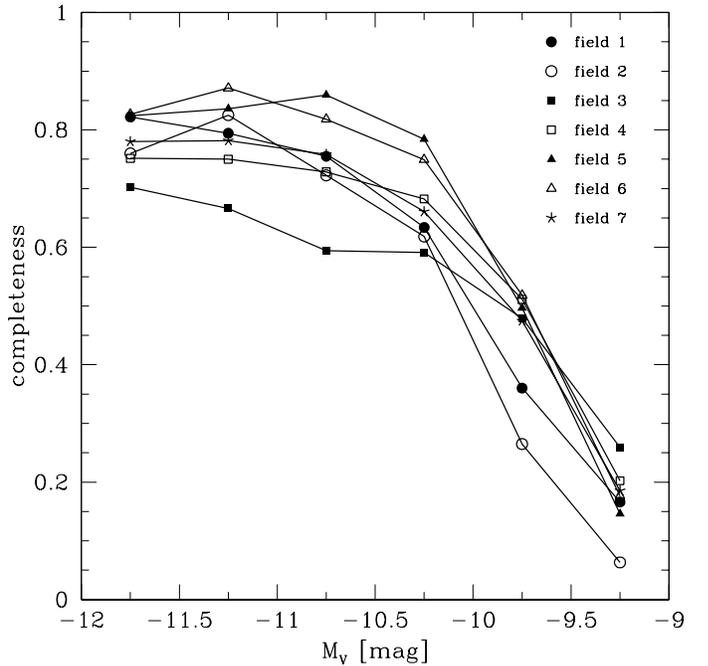}}
	\caption{Completeness as a function of magnitude (in 0.5 mag bins) for each of the seven cluster fields (cf. Fig.~\ref{fig:fields}).}
	\label{fig:completeness}
\end{figure}

In a first step, we made use of the SExtractor star/galaxy separator \citep{1996A&AS..117..393B} to sort out wrongly recovered foreground stars, requiring \texttt{class\_star} $<0.05$.  Aiming at the rejection of high surface brightness and barely resolved background objects, we then applied several cuts to other SExtractor output-parameters. The objects to be rejected were required to have an exponential scale length shorter than $1''$. This is close to the seeing limit of our images and it is the maximum scale length of the questionable objects from Sect.~\ref{sec:scalings}. The artificial galaxies with scale lengths $<1''$ define well localised areas in plots of \texttt{magbest} versus \texttt{mupeak}, \texttt{area} and \texttt{fwhm} (see Fig.~\ref{fig:artgal}). Since also some of the previously selected dwarf galaxy candidates scatter into the same areas, we finally rejected only those objects that \textit{simultaneously} occupied the locus of barely resolved galaxies in all three parameters \texttt{mupeak}, \texttt{area} and \texttt{fwhm}. In this way, we miss only one of the previously selected probable cluster dwarf galaxies but reject more than 50\% of objects with a scale length shorter than $1''$. In order to further optimise the rejection of obvious background objects we additionally applied global cuts at the upper limit of \texttt{mupeak} and the lower limit of \texttt{fwhm} (Fig.~\ref{fig:artgal}). 

Without the application of the cuts, SExtractor recovers 75-85\% of the simulated galaxies at $M_V\leq-12$ mag, which reflects the geometrical incompleteness caused by blending. Applying the cuts in \texttt{mupeak}, \texttt{area} and \texttt{fwhm} rejects $\sim25$\% more artificial galaxies at $M_V\leq-12$~mag. This fraction is consistent with the fraction of visually classified actual galaxies with $M_V>-12$~mag that are excluded by applying the same cuts (9 out of 36). Given that we include all visually classified galaxies into the LF, we scale the completeness values for $M_V>-12$~mag up by 25\%, so that they are consistent with the geometrical completeness at $M_V=-12$~mag (see Fig.~\ref{fig:completeness}).

\begin{figure}
	\resizebox{\hsize}{!}{\includegraphics{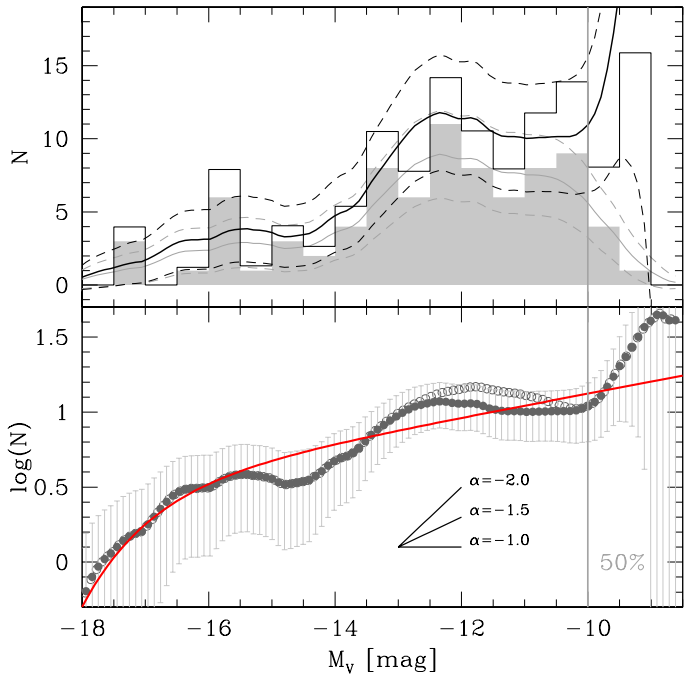}}
	\caption{Luminosity function of the Centaurus dwarf galaxies. The shaded histogram in the \textit{upper panel} shows the uncorrected galaxy number counts. The open histogram gives the completeness corrected number counts. The thin grey and thick black curves are binning independent representations of the counts (Epanechnikov kernel with 0.5 mag width). Dashed curves are the $1\sigma$ uncertainties. \textit{The lower panel} shows the completeness corrected galaxy number counts in logarithmic representation (filled circles). The best fitting Schechter function (red solid line) is overlaid. Open circles give the galaxy number counts including the questionable objects discussed in Sect.~\ref{sec:scalings}. Three different slopes $\alpha$ are indicated. The 50\% completeness limit (averaged over all fields) is given by the vertical line.}
	\label{fig:lumfunction}
\end{figure}

\begin{table}
	\caption{Fitting coefficients of the CMR, the magnitude-surface brightness relation, and the power-law slope $\alpha$ of the LF, with errors given in parentheses.}
	\label{tab:relations}
	\centering
	\begin{tabular}{lrrrrr}
	\hline\hline
	~ & \multicolumn{2}{c}{$(V-I)_0 = A\cdot M_V + B$} & \multicolumn{2}{c}{$\mu_{V,0} = C\cdot M_V +D$} \\ \cmidrule(r){2-3} \cmidrule(r){4-5}
	~ & $A$ & $B$ & $C$ & $D$ & $\alpha$ \\
	\hline
	Centaurus & $-0.042$  & $0.36$   & $0.57$   & $30.85$   & $-1.14$ \\
	~         & $(0.001)$ & $(0.02)$ & $(0.07)$ & $(0.87)$  & $(0.12)$ \\
	Hydra\,I  & $-0.044$  & $0.36$   & $0.67$   & $31.57$   & $-1.40$ \\
	~         & $(0.001)$ & $(0.01)$ & $(0.07)$ & $(0.99)$  & $(0.18)$ \\
	Fornax    & $-0.033$  & $0.52$   & $0.68$   & $32.32$   & $-1.33$ \\
	~         & $(0.004)$ & $(0.07)$ & $(0.04)$ & $(1.12)$  & $(0.08)$ \\
	\hline
	\end{tabular}
\end{table}

The completeness corrected galaxy luminosity function for $-17.5<M_V<-9.0$ mag is shown in Fig.~\ref{fig:lumfunction}. Due to the relatively low galaxy number counts (81 in this magnitude range), the LF is only moderately well represented by a \citet{1976ApJ...203..297S} function. From the best fitting Schechter function we derive a faint-end slope of $\alpha=-1.08\pm0.03$ (excluding galaxies fainter than $M_V=-10$ mag). As the slope $\alpha$ dominates the shape of the LF for magnitudes $M_V>-14$ mag, we alternatively fit a power-law to this interval, resulting in $\alpha=-1.14 \pm 0.12$. This characterises best the faint-end slope of the LF. Our result is consistent with the results of \citet{2006AJ....132..347C}, who found $\alpha \sim -1.4\pm0.2$ for the Centaurus cluster. They used statistical corrections as well as spectroscopic redshifts and surface brightness--magnitude criteria for the construction of the LF.

The nine questionable objects discussed in Sect.~\ref{sec:scalings} have absolute magnitudes of $-12.4<M_V<-11.2$ mag. Including them into the LF does not significantly change the slope $\alpha$ in the interval $-14<M_V<-10$ mag (see bottom panel of Fig.~\ref{fig:lumfunction}). Fitting a power-law leads to $\alpha=-1.17 \pm 0.12$.

In Table~\ref{tab:relations} the slope $\alpha$ is compared to the ones derived for the Fornax and Hydra\,I clusters \citep{2007A&A...463..503M, 2008A&A...486..697M}. Also for those clusters $\alpha$ is obtained by fitting a power-law to the faint end of the galaxy LF ($M_V>-14$ mag). With $-1.1\gtrsim \alpha \gtrsim -1.4$ all slopes are significantly shallower than the predicted slope of $\sim -2$ for the mass spectrum of cosmological dark-matter haloes \citep[e.g.][]{1974ApJ...187..425P, 1999ApJ...524L..19M, 2001MNRAS.321..372J}.

\subsection{Structural parameters from S\'ersic fits}
\label{sec:sersic}
\begin{figure*}
	\centering	
	\includegraphics[width=17cm]{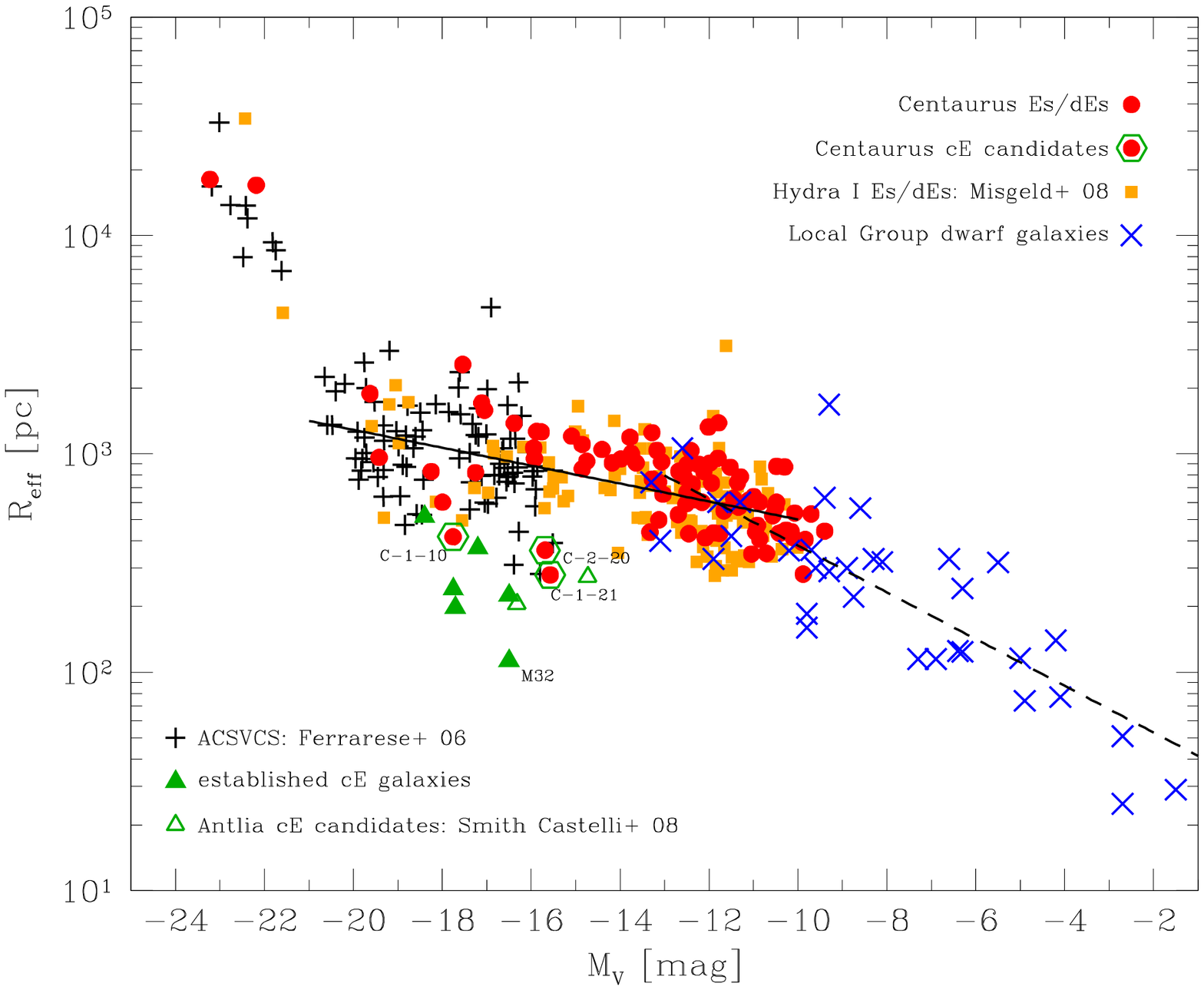}
	\caption{Plot of the effective radius $R_{\mathrm{eff}}$ against the absolute magnitude $M_V$ of the Centaurus early-type galaxies in comparison to Hydra\,I early-type galaxies from \citet{2008A&A...486..697M}, galaxies from the ACS Virgo Cluster Survey \citep{2006ApJS..164..334F},  Local Group dwarf galaxies and compact elliptical galaxies. The established cEs are M32 \citep{1992ApJ...399..462B, 1993ApJ...411..153B, 2003AJ....125.1926G}, A496cE \citep{2007A&A...466L..21C}, NGC 4486B \citep{2008arXiv0810.1681K}, NGC 5846A \citep{2005AJ....130.1502M, 2008MNRAS.tmp.1243S} and the two cEs from \citet{2005A&A...430L..25M}. The two Antlia cE candidates are taken from \citet{2008MNRAS.tmp.1243S}. Sources for the LG dwarf galaxies are: \citet{2003AJ....125.1926G} and \citet[][and references therein]{2007ApJ...663..948G} for Fornax, Leo I/II, Sculptor, Sextans, Carina and Ursa Minor; \citet{2008ApJ...684.1075M} for Draco, Canes Venatici I/II, Hercules, Leo IV, Coma Berenices, Segue I, Bootes I/II, Ursa Major~I/II and Willman~I; \citet{2006MNRAS.365.1263M} for And I/II/III/V/VI/VII and Cetus; \citet{2007ApJ...659L..21Z} for And IX/X; \citet{2006MNRAS.371.1983M} for And~XI/XII/XIII and \citet{2008ApJ...688.1009M} for And XVIII/XIX/XX. The solid line indicates the size--luminosity relation given by Eq.~(\ref{eq:size1}), the dashed line traces Eq.(\ref{eq:size2}).}
	\label{fig:sizelumidiag}
\end{figure*}

In addition to the exponential we also fitted S\'ersic models to the galaxy surface brightness profiles. The fit parameters central surface brightness $\mu_0$ and profile shape index $n$ are plotted versus the galaxy magnitude in Fig.~\ref{fig:sersic}. $\mu_0$ is given by $\mu_0=\mu_{\mathrm{eff}} - 2.5b_n/\ln(10)$, where $\mu_{\mathrm{eff}}$ is the effective surface brightness and $b_n$ is approximated by $b_n = 1.9992n - 0.3271$ for $0.5<n<10$ \citep[][and references therein]{2005PASA...22..118G}. Three bright cluster galaxies (C-4-03/NGC 4706, C-3-04 and C-7-07, see Table \ref{tab:Centaurussample}), morphologically classified as SAB(s)0, SB(s)0 and S0, showed two component surface brightness profiles (bulge + disk), which could not be fitted by a single S\'ersic profile. They were excluded from the analysis. 

The vast majority of cluster galaxies defines a continuous relation in the $\mu_0$ vs. $M_V$ diagram (top panel of Fig.~\ref{fig:sersic}). This relation runs from the faintest dwarf galaxies in our sample to bright cluster elliptical galaxies ($M_V\sim -20$ mag). Our results are consistent with other studies that report on a continuous relation for both dwarf galaxies and massive E/S0 galaxies \citep[e.g.][]{2003AJ....125.2936G, 2005A&A...430..411G, 2006ApJS..164..334F, 2008IAUS..246..377C, 2008A&A...486..697M}. Only the two brightest galaxies in our sample (NGC 4696 and NGC 4709) deviate from this relation.

The bottom panel of Fig.~\ref{fig:sersic} shows that also the profile shape index $n$ continuously rises with the galaxy magnitude for $M_V\lesssim-14$ mag. Only the brightest cluster galaxy, NGC 4696, has an exceptionally low S\'ersic index ($n=2.5$). For $M_V\gtrsim-14$ mag, $n$ basically stays constant with a mean value of 0.85. The spectroscopically confirmed background galaxies as well as the likely background objects in our sample can clearly be identified by their large S\'ersic index and their high central surface brightness in comparison to the cluster galaxies. This motivates again the rejection of those object from the cluster galaxy sample. Our results agree with former observations of a correlation of the S\'ersic index with the galaxy luminosity \citep[e.g.][]{1994MNRAS.268L..11Y, 2003Ap&SS.285...87I, 2006ApJS..164..334F, 2008A&A...486..697M}.

\begin{table*}
	\caption{Photometric and structural parameters of the cE galaxy candidates. References for radial velocities: $^{a}$\citep{1997A&A...327..952S}, $^{b}$\citep{2007ApJS..170...95C}.}
	\label{tab:cEs}
	\centering	
		\begin{tabular}{l c c c c c c c c c c c }
		\hline\hline
		ID & R.A. & Decl. & P.A. &$\varepsilon$  & $M_V$ & $(V-I)_0$ & $\mu_0$ & $R_{\mathrm{eff}}$ & $n$ & $v_{\mathrm{rad}}$ & $D_{\mathrm{NGC4696}}$ \\
		~ & (J2000.0) & (J2000.0) & [$\deg$] & & [mag] & [mag] & [mag/arcsec$^2$] & [pc] & ~ & [\kms] & [kpc] \\ 
		\hline
		C-1-10 & 12:48:53.9 & -41:19:05.3 & 69 & 0.1 & -17.76 & 1.18 & 17.42 & 418 & 1.23 & 2317$^{a}$ & 13 \\
		C-1-21 & 12:48:48.6 & -41:20:52.8 & 39 & $0.0\dots0.2$ & -15.57 & 1.18 & 18.56 & 279 & 1.27 & 3053$^{b}$ & 30 \\
		C-2-20 & 12:49:33.0 & -41:19:24.0 & -68 & 0.1 & -15.69 & 1.15 & 18.51 & 363 & 1.55 & -- & 109 \\
		\hline
		\end{tabular}
\end{table*}

\begin{figure*}
	\centering
	\subfigure{\includegraphics[width=6cm]{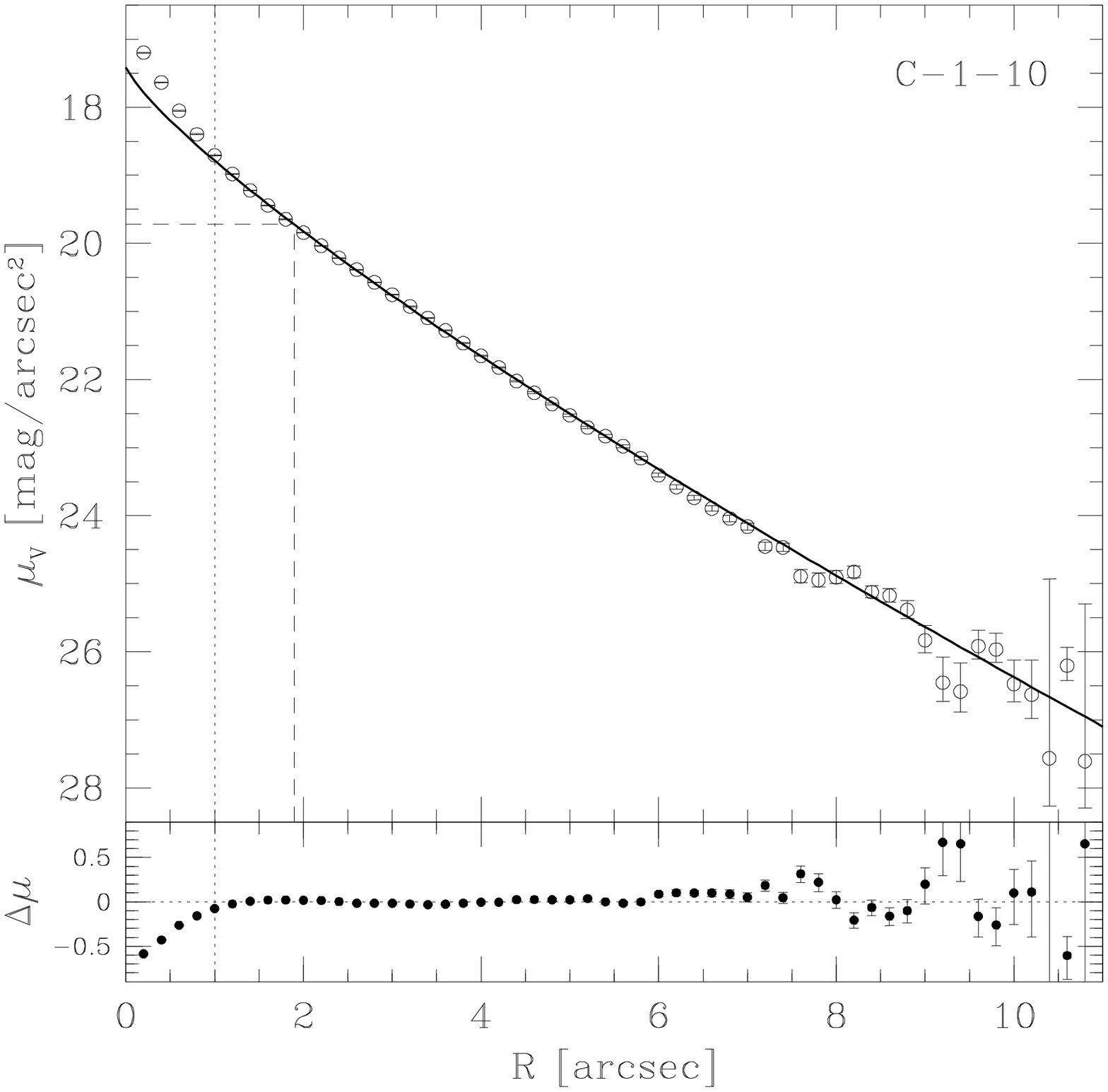}}
	\subfigure{\includegraphics[width=6cm]{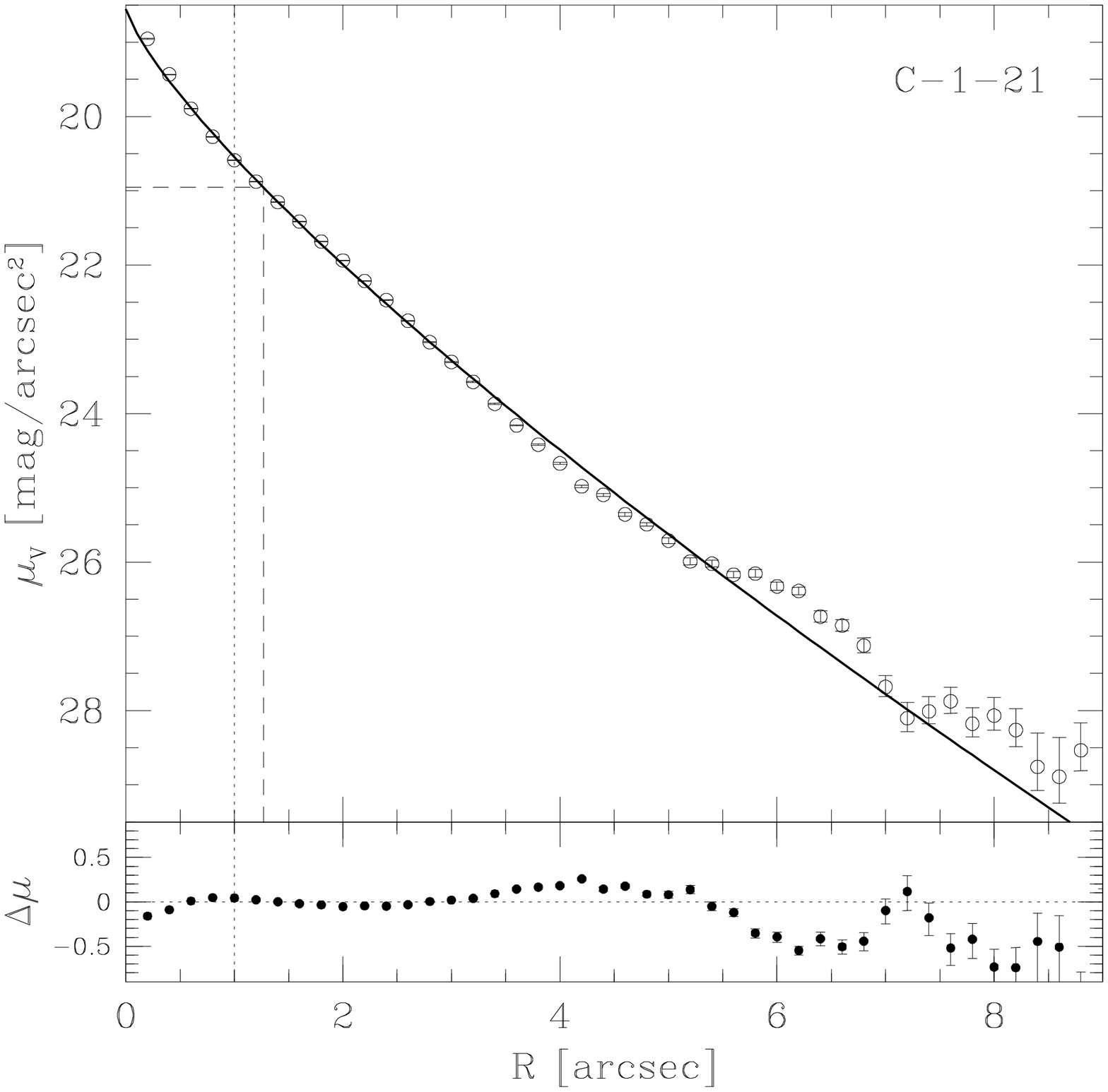}}
	\subfigure{\includegraphics[width=6cm]{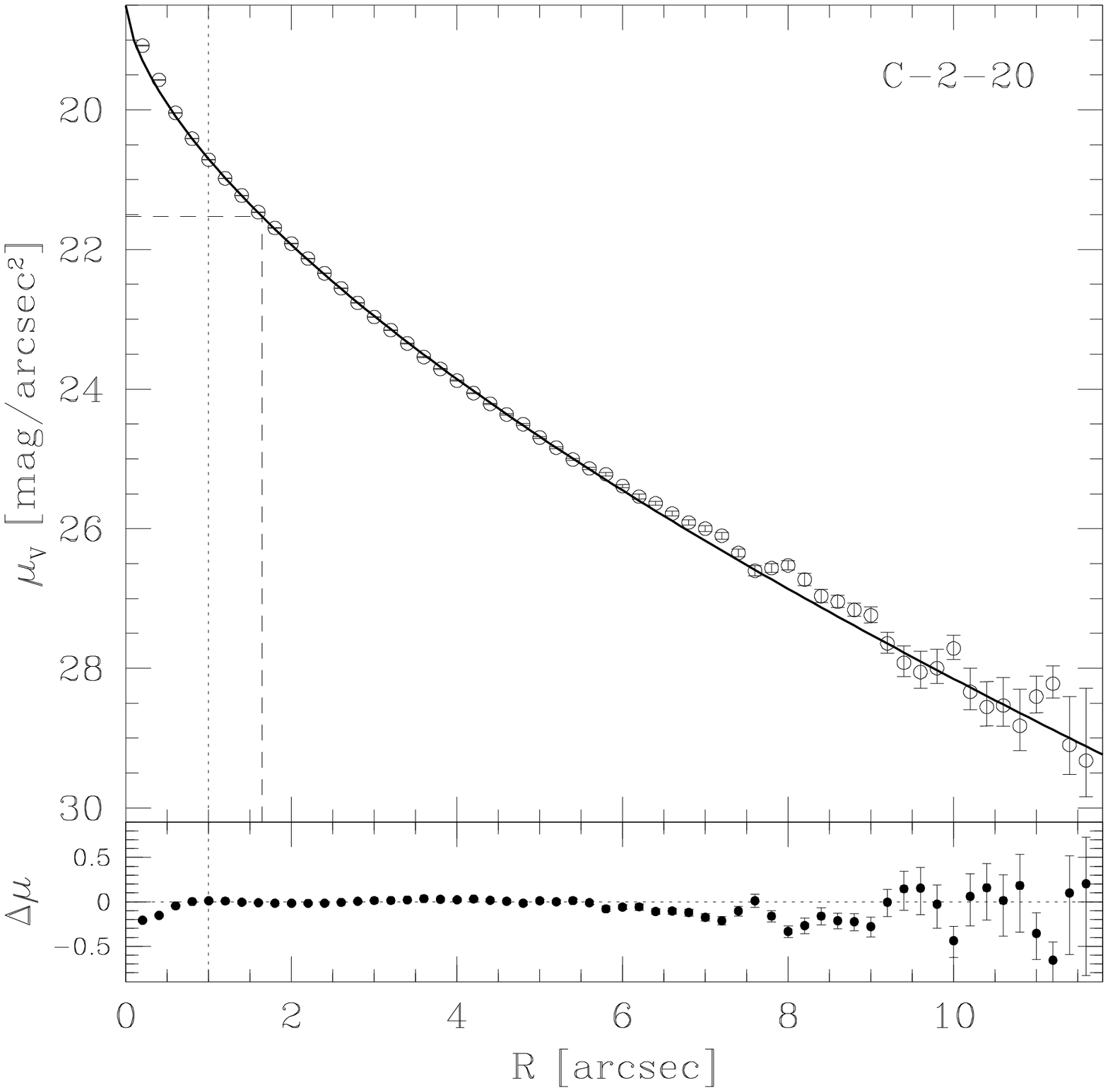}}
	\caption{Surface brightness profiles of the cE galaxy candidates. The extinction corrected surface brightness $\mu_V$ is plotted as a function of semi major axis radius $R$. The solid curve is the best fitting S\'ersic law. The residuals $\Delta\mu=\mu_{V,\mathrm{obs}} - \mu_{V,\mathrm{fit}}$ are shown in the lower panels. Vertical dotted lines mark the seeing affected region $R<1''$, which was excluded from the fit. The dashed lines indicate the effective surface brightness $\mu_{\mathrm{eff}}$ and the effective radius $R_{\mathrm{eff}}$.}
	\label{fig:profiles}
\end{figure*}

\subsubsection{Galaxy sizes}
In Fig.~\ref{fig:sizelumidiag} we show the effective radii and absolute magnitudes of the Centaurus early-type galaxies together with the Hydra\,I early-type galaxies from \citet{2008A&A...486..697M}, galaxies from the ACS Virgo Cluster Survey \citep{2006ApJS..164..334F}, Local Group dwarf galaxies and known compact elliptical galaxies. References for the LG dwarfs and the cEs are given in the caption. The sizes of the Centaurus galaxies agree very well with the sizes of the Hydra\,I galaxies and the sizes of bright galaxies ($-20<M_V<-16$~mag) in both samples are fully consistent with the ones obtained in \citet{2006ApJS..164..334F} for the ACS Virgo Cluster Survey galaxies. The apparent $g$-band magnitudes of the ACS Virgo Cluster Survey galaxies were transformed into absolute $V$-band magnitudes, using the transformation $V=g+0.026-0.307(g-z)$~mag given in \citet{2006ApJ...639..838P}. The transformation is derived from a study of diffuse star clusters around the ACS Virgo galaxies. Since the cluster colours are very similar to those of the host galaxies ($1.1<(g-z)<1.6$~mag), we consider the transformation a good approximation for the purposes of our study. We adopt a Virgo distance modulus of $(m-M)=31.09$~mag, corresponding to a scale of 80.1~pc/arcsec \citep{2007ApJ...655..144M}. For the calculation of the $V$-band magnitude of the cE galaxy NGC 5846A we applied $V-R=0.61$~mag \citep{1995PASP..107..945F} and the distance modulus $(m-M)=32.08$~mag, corresponding to scale of 126~pc/arcsec \citep[][and references therein]{2008MNRAS.tmp.1243S}.

The most striking feature in Fig.~\ref{fig:sizelumidiag} is the continuous size--luminosity relation over a wide magnitude range. The effective radius slowly increases as a function of galaxy magnitude for $-21<M_V<-10$~mag. The relation between $\log(R_{\mathrm{eff}})$ and $M_V$ is indicated by the solid line in Fig.~\ref{fig:sizelumidiag} and can be quantified as
\begin{equation}
\label{eq:size1}
 \log(R_{\mathrm{eff}})= -0.041(\pm 0.004) \cdot M_V + 2.29(\pm 0.06)
\end{equation}
with an rms of 0.17. At magnitudes fainter than $M_V\gtrsim -13$~mag, the slope of the relation becomes slightly steeper. A fit to the data yields
\begin{equation}
\label{eq:size2}
 \log(R_{\mathrm{eff}})= -0.107(\pm 0.007) \cdot M_V + 1.51(\pm 0.07)
\end{equation}
with an rms of 0.17 (dashed line in Fig.~\ref{fig:sizelumidiag}). In their study of photometric scaling relations of early-type galaxies in Fornax, Coma, Antlia, Perseus and the LG, \citet{2008arXiv0811.3198D} reported on a very similar behaviour. However, comparatively few data points are available for $M_V>-10$~mag, i.e. the regime of faint LG dwarf spheroidals, and there might be a bias towards the selection of more compact objects at fainter magnitudes, in the sense that at a given magnitude very extended low surface brightness galaxies are more likely to be missed than more compact ones. Moreover, the two smallest LG dwarf galaxies Segue\,I and Willman\,I \citep{2008ApJ...684.1075M} are suspected to be globular star clusters or dSphs out of dynamical equilibrium, close to disruption, rather than ordinary dwarf galaxies \citep{2007ApJ...663..948G}.

Two groups of objects clearly deviate from the size--luminosity relations defined by the other objects. These are the brightest core galaxies ($M_V\lesssim -21$~mag) which show a very strong dependence of effective radius on absolute magnitude, and a few rather compact galaxies which fall below the main body of normal elliptical galaxies. The latter are discussed in more detail in the following subsection.

\subsubsection{Compact elliptical galaxy candidates}
\label{sec:cEs}
Three unusual objects, having rather small effective radii compared to other cluster galaxies with similar magnitudes, stand out in Fig.~\ref{fig:sizelumidiag}. Do they belong to the class of the so-called compact elliptical galaxies (cEs)? For the three candidates, Table \ref{tab:cEs} lists the coordinates, the absolute magnitude $M_V$, the extinction corrected colour $(V-I)_0$, the central surface brightness $\mu_0$, the effective radius $R_{\mathrm{eff}}$, the S\'ersic index $n$, the available radial velocity $v_{\mathrm{rad}}$ and the projected distance $D_{\mathrm{NGC4696}}$ to the central cluster galaxy NGC~4696. Also given are the position angle (P.A.) and the ellipticity $\varepsilon$ used for the fit of elliptical isophotes to the galaxy image. In Fig.~\ref{fig:profiles} we show for each of the cE galaxy candidates the S\'ersic fits and the according residuals to their surface brightness profiles. In the following three paragraphs we describe in detail how the photometric parameters were obtained and try to judge whether the objects belong to the class of cE galaxies.

\paragraph{C-1-10}
is a spectroscopically confirmed member of the Centaurus cluster. It is listed as CCC 70 in the Centaurus Cluster Catalogue and it is morphologically classified as an E0(M32) galaxy \citep{1997A&AS..124....1J}. The isophote fitting was performed on the 30s exposure, since the long-exposure image was saturated at the object centre. Due to the projected proximity of C-1-10 to the giant galaxy NGC 4696, we created and subtracted a model of the latter before modelling the dwarf galaxy.

With an effective radius of $1.90''$ (418 pc) C-1-10 is the most compact object among the galaxies with similar magnitude in our sample. However, it is larger than most of the cE galaxies mentioned in the literature (see Fig.~\ref{fig:sizelumidiag}). Only for NGC 5846A an even larger effective radius of $\sim500$ pc is reported \citep{2005AJ....130.1502M}. Moreover, C-1-10 does not have a particular high central surface brightness, but it falls exactly on the sequence of regular cluster dwarf galaxies (see upper panel of Fig.~\ref{fig:sersic}). Also its colour is consistent with the cluster CMR (Fig.~\ref{fig:cmd}). Given these properties, C-1-10 is rather a small elliptical galaxy than an exemplar of a cE galaxy.

\paragraph{C-1-21}
is a confirmed Centaurus member \citep{2007ApJS..170...95C}. The best model for C-1-21 was obtained with fixed centre coordinates and position angle, while the ellipticity $\varepsilon$ was allowed to vary ($0.0<\varepsilon<0.2$). Its effective radius of $1.27''$, or 279 pc, is at least three times smaller than the ones of other cluster galaxies of the same luminosity (Fig.~\ref{fig:sizelumidiag}). This is comparable to the size of the two M32 twins in Abell 1689 \citep{2005A&A...430L..25M}, the cE galaxy A496cE \citep{2007A&A...466L..21C} and the cE candidate FS90~192 in the Antlia cluster \citep{2008MNRAS.tmp.1243S}. The central surface brightness of C-1-21 is about 2~mag/arcsec$^2$ higher than the one of equally bright cluster galaxies (see Fig.~\ref{fig:sersic}) and its colour is about 0.15~mag redder than expected from the cluster CMR (Eq.~(\ref{eq:cmr})). Interestingly, both colour and central surface brightness would be consistent with other cluster galaxies, if the object was about 2~mag brighter. This suggests that C-1-21 might originate from a higher mass elliptical or spiral galaxy, which was stripped by the strong tidal field of NGC 4696 \citep[e.g.][]{1973ApJ...179..423F, 2001ApJ...557L..39B}.

Since three common characteristics of cE galaxies, namely the small effective radius, the high central surface brightness and the projected location close to a brighter galaxy, are given, we consider C-1-21 a true cE galaxy.

\paragraph{C-2-20} has no available spectroscopic redshift. Whether it is a cluster galaxy or a background galaxy can at this point only be determined by an educated guess on the basis of its morphological and photometric properties. Absolute magnitude, colour and central surface brightness are very similar to those of C-1-21 (see Table \ref{tab:cEs}). Its very compact morphology ($R_{\mathrm{eff}}=363$ pc) suggests that it is indeed a cluster cE. However, its relatively isolated position (see Fig.~\ref{fig:fields}), far away from the giant galaxies, is unusual for a cE galaxy. We conclude that C-2-20 is definitely a good candidate for a cE galaxy, but whether it really belongs to the Centaurus cluster has to be confirmed by spectroscopic redshift measurements.

\section{Summary and discussion}
\label{sec:discussion}
Based on deep VLT/FORS1 imaging data in Johnson $V$ and $I$ we studied the early-type dwarf galaxy population of the Centaurus cluster. We combined visual classification and SExtractor based detection routines in order to select candidate objects on the images (Sect.~\ref{sec:sample}).

We investigated fundamental scaling relations, such as the colour--magnitude relation and the magnitude-surface brightness relation (Sect.~\ref{sec:scalings}). Both relations were found to be consistent with the ones in the Fornax and Hydra\,I galaxy clusters (see Table~\ref{tab:relations}). Moreover, LG dwarf galaxies projected to the Centaurus distance follow the same magnitude-surface brightness relation. Both scaling relations enabled us to define a sample of probable cluster galaxies, which was used to construct the galaxy luminosity function down to a limiting magnitude of $M_V=-10$ mag (Sect.~\ref{sec:lumfunction}).

\subsection{The faint end of the galaxy LF}
From the completeness corrected galaxy number counts we derive a very flat faint-end slope of the Centaurus galaxy LF. A power law describes best the shape of the faint end of the LF. We measure a slope of $\alpha = -1.14 \pm 0.12$ (see Fig.~\ref{fig:lumfunction} and Table~\ref{tab:relations}). A similar value is obtained when fitting a Schechter function to the data ($\alpha \sim -1.1$). A flat LF for the Centaurus cluster was also derived by \citet{2006AJ....132..347C}. Moreover, our result is consistent with the flat LFs observed in other nearby galaxy clusters, for which the LF was similarly constructed using morphological selection criteria \citep[e.g.][]{2002MNRAS.335..712T, 2003A&A...397L...9H, 2007A&A...463..503M, 2008A&A...486..697M}. 

The cluster membership assignment by means of morphology and surface brightness is, of course, the key step for the entire analysis. Misclassifications have to be prevented as far as possible, but they can hardly be avoided entirely. In particular, it is often difficult to distinguish cluster dwarf galaxies with rather high surface brightnesses from background galaxies which only resemble the cluster galaxies. We indeed identified nine questionable objects in our sample, having colours similar to the cluster dwarfs but with significantly higher surface brightnesses along with a rather compact morphology (cf. Sect.~\ref{sec:scalings}). Due to the fact that they deviate more than $2\sigma$ from the magnitude-surface brightness relation (Eq. (\ref{eq:magmu})), which is defined by the probable cluster dwarf galaxies (see Fig.~\ref{fig:magmu}), we conclude that most of those objects do not belong to the cluster. In any case, they do not significantly influence the galaxy LF, raising the faint-end slope marginally from $\alpha=-1.14$ to $\alpha=-1.17$ (cf. Sect.~\ref{sec:lumfunction}).

Another caveat of a morphological selection is that one could potentially misclassify compact, M32-like cluster members as background objects, or vice versa \citep{2002MNRAS.335..712T}. We found three cE galaxy candidates in our sample, of which two are confirmed cluster members and the third one has photometric properties very similar to confirmed cluster galaxies (see Sect.~\ref{sec:cEs}). However, cE galaxies are rare and have (like our candidates) rather bright magnitudes (Fig.~\ref{fig:sizelumidiag}), so that they do not affect the shape of the faint end of the galaxy LF.

Altogether, we are confident not to have misclassified a large number of objects, since we were sensitive to very low surface brightnesses, the seeing of our images was excellent, and we made use of photometric scaling relations to substantiate the morphological classifications.

Besides Virgo, Fornax and Hydra\,I, Centaurus is now the fourth galaxy cluster in the local Universe, whose LF has been investigated down to the regime of dwarf spheroidal galaxies ($M_V\sim -10$ mag). Flat luminosity functions, which are contradictory to the predicted mass spectrum of cosmological dark-matter haloes \citep[e.g.][]{1974ApJ...187..425P, 1999ApJ...524L..19M, 2001MNRAS.321..372J}, have been derived for all of these environments. It seems to become apparent that this discrepancy to hierarchical cold dark matter models of galaxy formation is a common feature of various galaxy clusters/groups. However, one has to note that we primarily investigate early-type galaxies in a rather limited region in the central part of the cluster. The slope of the LF might considerably change when including the outer parts of the cluster into the analysis. Moreover, although not found in our study, late-type dwarf irregular galaxies might affect the shape of the LF as well.

In the end it is essential to have \textit{direct} cluster membership determination via deep spectroscopic surveys for magnitudes $M_V\gtrsim -14$ mag, where the faint-end slope $\alpha$ starts to dominate the shape of the LF (cf. Fig.~\ref{fig:lumfunction}). Beyond the Local Group, this has up to now only been achieved in studies of the galaxy clusters Fornax, Perseus and Virgo \citep[e.g.][]{1999A&AS..134...75H, 2001ApJ...548L.139D, 2008MNRAS.383..247P, 2008AJ....135.1837R}. The next step should therefore be the extension of those surveys to other galaxy clusters like Centaurus or Hydra\,I in order to thoroughly verify the results of the photometric studies. With a reasonable amount of observing time ($\sim 2$ hours integration time) the spectroscopic confirmation of low surface brightness objects is technically feasible for objects with $\mu_V \lesssim 25$~mag/arcsec$^2$, using low-resolution spectrographs like FORS or VIMOS at the VLT (see \texttt{http://www.eso.org/observing/etc/} for exposure time calculators). This surface brightness limit corresponds to an absolute magnitude limit of $M_V\sim -11$ mag at the distance of Centaurus (cf. Fig.~\ref{fig:magmu}). Naturally, at a given magnitude the surface brightness limit introduces a bias towards more successfully measuring the redshifts of smaller objects with higher surface brightnesses. However, the cluster membership assignment via morphological classification for exactly those rather compact objects turns out to be more difficult than for extended low surface brightness galaxies (see Sect.~\ref{sec:scalings}). This means that the observational bias primarily excludes objects for which the morphological membership assignment is more accurate, thus, contamination by background objects is smaller.

\subsection{The dependency of effective radius on luminosity}
We derived structural parameters, such as central surface brightness $\mu_0$, effective radius $R_{\mathrm{eff}}$ and profile shape index $n$, of the probable cluster galaxies by fitting \citet{1968adga.book.....S} models to the galaxy surface brightness profiles (Sect.~\ref{sec:sersic}).

In plots of $\mu_0$ and $n$ versus the galaxy magnitude we observe continuous relations, ranging 10 orders of magnitude from $M_V=-20$ mag to the magnitude limit of our survey (Fig.~\ref{fig:sersic}). This confirms observations of continuous relations in the LG and other galaxy clusters, such as Fornax, Virgo and Hydra\,I \citep[e.g.][]{1994MNRAS.268L..11Y, 2003AJ....125.2936G, 2006ApJS..164..334F, 2008IAUS..246..377C, 2008A&A...486..697M}. Only the brightest cluster galaxies have central surface brightnesses which are lower than expected from the extrapolation of the relation defined by galaxies of lower luminosity. The deviation of these core galaxies from the $M_V$--$\mu_0$ relation can be explained by mass depleted central regions due to the dynamical interaction of a supermassive black hole binary \citep[][and references therein]{2006ApJS..164..334F}. A different point of view is, however, that these galaxies belong to a different sequence, almost perpendicular to the dE sequence, populated with bright early-type galaxies ($M_V\lesssim-20$ mag), for which the surface brightness decreases with increasing magnitude \citep[e.g.][]{1985ApJ...295...73K, 1992ApJ...399..462B, 2008arXiv0810.1681K}.

The size--luminosity diagram is another tool to visualise a dis-/continuity between dwarf and giant elliptical galaxies. Combining our data with studies of early-type galaxies in Virgo \citep{2006ApJS..164..334F}, Hydra\,I \citep{2008A&A...486..697M} and the LG, we find a well defined sequence in such a diagram (see Fig.~\ref{fig:sizelumidiag}). For a wide magnitude range ($-21\lesssim M_V \lesssim -13$~mag) the effective radius changes little with luminosity. For fainter magnitudes the slope of the size--luminosity relation steepens and the sequence continues all the way down to the ultra-faint LG dwarf galaxies ($M_V\sim -4$~mag), which have been identified in the SDSS \citep[e.g.][]{2006MNRAS.371.1983M, 2008ApJ...684.1075M, 2006MNRAS.365.1263M, 2007ApJ...663..948G, 2007ApJ...659L..21Z, 2008ApJ...688.1009M}. Only the brightest core galaxies and compact elliptical galaxies deviate from the relation of ordinary elliptical galaxies. Both the continuous surface brightness vs. absolute magnitude relation and the continuous sequence in the size--luminosity diagram are consistent with the interpretation that dwarf galaxies and more massive elliptical galaxies are one family of objects. In this scenario the scaling relations are caused by the gradual change of the S\'ersic index $n$ with the galaxy magnitude \citep[e.g.][]{1997ASPC..116..239J, 2003AJ....125.2936G, 2005A&A...430..411G}.

In contrast to this interpretation, \citet{2008arXiv0810.1681K} and \citet{2008ApJ...689L..25J} most recently reported on a pronounced dichotomy of elliptical and spheroidal galaxies in the size--luminosity diagram, which is \textit{not} caused by the gradual change of the galaxy light profile with luminosity. \citet{2008arXiv0810.1681K} reaffirm results of older studies \citep[e.g.][]{1985ApJ...295...73K,1991A&A...252...27B, 1992ApJ...399..462B, 1993ApJ...411..153B} and claim that the dwarf galaxy sequence intersects at $M_V\sim-18$~mag a second (much steeper) sequence, which consists of giant elliptical and S0 galaxies and extends to the regime of cE galaxies \citep[see also][]{2008MNRAS.386..864D}. They conclude that massive elliptical and spheroidal galaxies are physically different and have undergone different formation processes. The latter were created by the transformation of late-type galaxies into spheroidals, whereas the giant ellipticals formed by mergers. By comparing the observations to models of ram-pressure stripping and galaxy harassment, \citet{2008A&A...489.1015B} indeed find evidence for different formation mechanisms. Although the bulk of galaxies investigated in \citet{2006ApJS..164..334F} falls into the magnitude range where the dichotomy should become apparent, these authors did not report on two distinct sequences \citep[but see appendix~B in][]{2008arXiv0810.1681K}. 

Based on our data we cannot confirm the existence of an E -- Sph dichotomy. Our data rather show a continuous sequence of structural properties across a wide range of galaxy luminosities (masses), supporting the interpretation that dSphs as well as Es are one family of objects. Merely, the most massive core galaxies and the cE galaxies are clearly separated from normal ellipticals. In this context, however, one has to keep in mind that cE galaxies are extremely rare, whereas normal elliptical galaxies are much more frequent. This raises the question of which process is responsible for the peculiar properties of cEs. More than one formation channel is being discussed for cEs like M32. If they are the results of galaxy threshing \citep{2001ApJ...557L..39B}, their ``original'' location in Fig.~\ref{fig:sizelumidiag} was at higher luminosity and larger radius, probably consistent with the main body of ordinary elliptical galaxies. They would rightfully deserve to be termed low mass counterparts of giant ellipticals \citep[][and references therein]{2008arXiv0810.1681K} only if they were intrinsically compact at the time of their formation \citep{2002AJ....124..310C}.

\bibliographystyle{aa}
\bibliography{centaurusdwarfs}

\Online

\begin{appendix}
\section{Tables}
Table \ref{tab:photcal} gives the photometric calibration coefficients for the seven observed cluster fields, as indicated in Fig. \ref{fig:fields}, and the background field (field 8). Zero points (ZP), extinction coefficients $k$ and colour terms (CT) are given for the two filters $V$ and $I$.

\bigskip
\noindent
Table \ref{tab:Centaurussample} lists the photometric parameters of our sample of 92 probable Centaurus cluster early-type galaxies. The table is ordered by increasing apparent magnitude. The first column gives the object ID, in which the first number refers to the field in which the object is located (cf. Fig.~\ref{fig:fields}). Right ascension and declination (J2000.0) are given in columns two and three. The fourth and fifth column contain the extinction corrected magnitude $V_0$ and colour $(V-I)_0$. In columns six and seven, the central surface brightness $\mu_{V,0}$ and the scale length $h_R$ of an exponential fit to the surface brightness profile are listed. $\mu_{V,0}$ and $h_R$ are not given for objects, whose surface brightness profiles are not well described by an exponential law, i.e. objects brighter than $V_0=16.1$ mag and the three cE galaxy candidates (see Sect.~\ref{sec:cEs}). Columns eight, nine and ten give the effective surface brightness $\mu_{\mathrm{eff}}$, the effective radius $R_{\mathrm{eff}}$ and the profile shape index $n$, as obtained from a S\'ersic fit. The physical scale is 0.22 kpc/arcsec at the assumed distance modulus of $(m-M)=33.28$ mag \citep{2005A&A...438..103M}.

\begin{table}[hb!]
	\caption{Photometric calibration coefficients.}
	\label{tab:photcal}
	\centering	
		\begin{tabular}{l r r r r r r}
		\hline\hline
		Field & $\mathrm{ZP}_V$ & $\mathrm{ZP}_I$ & $k_V$ & $k_I$ & $\mathrm{CT}_V$ & $\mathrm{CT}_I$ \\
		\hline
		1 & 27.531 & 26.672 & -0.135 & -0.061 & 0.026 & -0.061 \\
		2 & 27.523 & 26.675 & -0.135 & -0.061 & 0.026 & -0.061 \\
		3 & 27.518 & 26.668 & -0.135 & -0.061 & 0.026 & -0.061 \\
		4 & 27.505 & 26.659 & -0.135 & -0.061 & 0.026 & -0.061 \\
		5 & 27.516 & 26.678 & -0.135 & -0.061 & 0.026 & -0.061 \\
		6 & 27.516 & 26.678 & -0.135 & -0.061 & 0.026 & -0.061 \\
		7 & 27.531 & 26.672 & -0.135 & -0.061 & 0.026 & -0.061 \\
		8 & 27.510 & 26.628 & -0.135 & -0.061 & 0.026 & -0.061 \\
		\hline
		\end{tabular}
\end{table}

\clearpage
\onecolumn

{\tiny
\longtab{2}{
\begin{longtable}{lccccccccc}
\caption{\label{tab:Centaurussample} Catalogue of the probable Centaurus cluster early-type galaxies in our sample.}\\
\hline\hline
ID & $\alpha$(2000.0) & $\delta$(2000.0) & $V_0$ & $(V-I)_0$ & $\mu_{V,0}$ & $h_R$ & $\mu_{\mathrm{eff}}$ & $R_{\mathrm{eff}}$ & $n$ \\
~ & [h:m:s] & [$^\circ$:$\arcmin$:$\arcsec$] & [mag] & [mag] & [mag arcsec$^{-2}$] & [arcsec] & [mag arcsec$^{-2}$] & [arcsec] & ~ \\
\hline
\endfirsthead
\caption{continued.}\\
\hline\hline
ID & $\alpha$(2000.0) & $\delta$(2000.0) & $V_0$ & $(V-I)_0$ & $\mu_{V,0}$ & $h_R$ & $\mu_{\mathrm{eff}}$ & $R_{\mathrm{eff}}$ & $n$ \\
~ & [h:m:s] & [$^\circ$:$\arcmin$:$\arcsec$] & [mag] & [mag] & [mag arcsec$^{-2}$] & [arcsec] & [mag arcsec$^{-2}$] & [arcsec] & ~ \\
\hline
\endhead
\hline
\endfoot

C-1-01$^c$ & 12:48:49.3 & $-$41:18:39.1 & $10.05 \pm 0.01$ & $1.25 \pm 0.03$ &  &  & $22.29 \pm 0.01$ & $82.07 \pm 0.03$ & $2.49 \pm 0.01$ \\ 
C-3-02$^d$ & 12:50:04.0 & $-$41:22:54.1 & $11.10 \pm 0.02$ & $1.28 \pm 0.02$ &  &  & $23.28 \pm 0.01$ & $77.25 \pm 0.14$ & $4.02 \pm 0.01$ \\ 
C-4-03$^{b,e}$ & 12:49:54.2 & $-$41:16:44.9 & $12.56 \pm 0.01$ & $1.22 \pm 0.01$ &  &  &  &  &  \\ 
C-3-04$^b$ & 12:49:38.0 & $-$41:23:20.2 & $13.17 \pm 0.01$ & $1.13 \pm 0.01$ &  &  &  &  &  \\ 
C-4-05 & 12:49:51.6 & $-$41:13:34.2 & $13.65 \pm 0.01$ & $1.17 \pm 0.02$ &  &  & $21.29 \pm 0.01$ & $8.57 \pm 0.01$ & $3.57 \pm 0.01$ \\ 
C-3-06 & 12:50:08.0 & $-$41:23:49.3 & $13.86 \pm 0.01$ & $1.21 \pm 0.01$ &  &  & $19.79 \pm 0.01$ & $4.38 \pm 0.01$ & $2.61 \pm 0.01$ \\ 
C-7-07$^b$ & 12:48:43.5 & $-$41:38:36.8 & $14.05 \pm 0.01$ & $1.11 \pm 0.01$ &  &  &  &  &  \\ 
C-1-08 & 12:48:31.1 & $-$41:18:23.3 & $15.03 \pm 0.03$ & $1.12 \pm 0.04$ &  &  & $20.52 \pm 0.01$ & $3.77 \pm 0.01$ & $2.09 \pm 0.01$ \\ 
C-2-09 & 12:49:18.6 & $-$41:20:07.3 & $15.29 \pm 0.01$ & $1.13 \pm 0.01$ &  &  & $20.22 \pm 0.01$ & $2.73 \pm 0.01$ & $1.41 \pm 0.01$ \\ 
C-1-10 & 12:48:53.9 & $-$41:19:05.3 & $15.52 \pm 0.01$ & $1.18 \pm 0.01$ &  &  & $19.72 \pm 0.01$ & $1.90 \pm 0.01$ & $1.23 \pm 0.01$ \\ 
C-3-11 & 12:49:40.2 & $-$41:21:60.0 & $15.74 \pm 0.02$ & $1.02 \pm 0.03$ &  &  & $23.54 \pm 0.01$ & $11.67 \pm 0.01$ & $1.38 \pm 0.01$ \\ 
C-4-12 & 12:49:42.0 & $-$41:13:44.9 & $16.03 \pm 0.01$ & $1.10 \pm 0.01$ &  &  & $21.75 \pm 0.01$ & $3.73 \pm 0.01$ & $1.71 \pm 0.01$ \\ 
C-4-13 & 12:49:56.4 & $-$41:15:35.8 & $16.17 \pm 0.01$ & $1.10 \pm 0.01$ & $21.24 \pm 0.01$ & $4.62 \pm 0.01$ & $23.08 \pm 0.01$ & $7.77 \pm 0.01$ & $1.08 \pm 0.01$ \\ 
C-1-14 & 12:48:39.8 & $-$41:16:05.7 & $16.23 \pm 0.01$ & $1.09 \pm 0.01$ & $21.26 \pm 0.01$ & $4.24 \pm 0.01$ & $23.12 \pm 0.01$ & $7.19 \pm 0.01$ & $1.15 \pm 0.01$ \\ 
C-5-15 & 12:48:36.1 & $-$41:26:23.2 & $16.89 \pm 0.01$ & $0.89 \pm 0.02$ & $21.42 \pm 0.01$ & $3.72 \pm 0.01$ & $23.31 \pm 0.01$ & $6.28 \pm 0.01$ & $1.32 \pm 0.01$ \\ 
C-1-16 & 12:48:30.1 & $-$41:19:17.8 & $17.34 \pm 0.01$ & $0.96 \pm 0.01$ & $21.36 \pm 0.01$ & $2.80 \pm 0.01$ & $23.29 \pm 0.01$ & $4.85 \pm 0.01$ & $1.15 \pm 0.01$ \\ 
C-2-17 & 12:49:02.0 & $-$41:15:33.7 & $17.35 \pm 0.01$ & $0.96 \pm 0.01$ & $21.24 \pm 0.01$ & $2.67 \pm 0.01$ & $22.97 \pm 0.01$ & $4.33 \pm 0.01$ & $0.89 \pm 0.01$ \\ 
C-3-18 & 12:49:56.2 & $-$41:24:04.0 & $17.40 \pm 0.01$ & $1.00 \pm 0.01$ & $21.97 \pm 0.01$ & $4.18 \pm 0.01$ & $23.33 \pm 0.01$ & $5.75 \pm 0.01$ & $1.43 \pm 0.01$ \\ 
C-3-19 & 12:49:54.1 & $-$41:20:21.9 & $17.50 \pm 0.01$ & $1.05 \pm 0.01$ & $21.53 \pm 0.01$ & $3.39 \pm 0.01$ & $23.40 \pm 0.01$ & $5.73 \pm 0.01$ & $1.11 \pm 0.01$ \\ 
C-2-20 & 12:49:33.0 & $-$41:19:24.0 & $17.59 \pm 0.01$ & $1.15 \pm 0.01$ &  &  & $21.53 \pm 0.01$ & $1.65 \pm 0.01$ & $1.55 \pm 0.01$ \\ 
C-1-21 & 12:48:48.6 & $-$41:20:52.8 & $17.71 \pm 0.01$ & $1.18 \pm 0.01$ &  &  & $20.95 \pm 0.01$ & $1.27 \pm 0.01$ & $1.27 \pm 0.01$ \\ 
C-2-22 & 12:49:22.7 & $-$41:15:18.7 & $18.19 \pm 0.02$ & $0.95 \pm 0.04$ & $22.38 \pm 0.01$ & $3.63 \pm 0.01$ & $23.98 \pm 0.01$ & $5.48 \pm 0.01$ & $1.39 \pm 0.01$ \\ 
C-3-23 & 12:49:46.6 & $-$41:22:08.7 & $18.42 \pm 0.01$ & $0.97 \pm 0.01$ & $22.51 \pm 0.01$ & $3.00 \pm 0.01$ & $24.33 \pm 0.01$ & $5.02 \pm 0.01$ & $1.04 \pm 0.01$ \\ 
C-2-24 & 12:49:05.4 & $-$41:18:25.6 & $18.42 \pm 0.01$ & $0.87 \pm 0.02$ & $22.15 \pm 0.01$ & $2.40 \pm 0.01$ & $23.88 \pm 0.01$ & $3.88 \pm 0.01$ & $0.87 \pm 0.01$ \\ 
C-2-25 & 12:49:32.3 & $-$41:20:23.8 & $18.53 \pm 0.01$ & $0.99 \pm 0.02$ & $22.38 \pm 0.01$ & $2.58 \pm 0.02$ & $24.16 \pm 0.01$ & $4.21 \pm 0.03$ & $0.99 \pm 0.01$ \\ 
C-4-26 & 12:49:48.7 & $-$41:14:18.3 & $18.87 \pm 0.01$ & $0.88 \pm 0.02$ & $22.72 \pm 0.01$ & $2.78 \pm 0.01$ & $24.60 \pm 0.01$ & $4.76 \pm 0.02$ & $1.09 \pm 0.01$ \\ 
C-2-27 & 12:49:06.5 & $-$41:16:13.4 & $19.10 \pm 0.01$ & $0.96 \pm 0.02$ & $22.88 \pm 0.01$ & $2.47 \pm 0.01$ & $24.70 \pm 0.01$ & $4.13 \pm 0.03$ & $1.05 \pm 0.01$ \\ 
C-3-28 & 12:49:56.2 & $-$41:23:23.4 & $19.29 \pm 0.02$ & $0.86 \pm 0.03$ & $22.84 \pm 0.01$ & $2.59 \pm 0.01$ & $24.60 \pm 0.01$ & $4.31 \pm 0.01$ & $0.73 \pm 0.01$ \\ 
C-4-29 & 12:49:56.1 & $-$41:16:56.4 & $19.50 \pm 0.04$ & $0.87 \pm 0.05$ & $23.32 \pm 0.01$ & $3.18 \pm 0.01$ & $25.18 \pm 0.02$ & $5.41 \pm 0.05$ & $1.06 \pm 0.02$ \\ 
C-3-30$^a$ & 12:49:34.1 & $-$41:22:38.8 & $19.53 \pm 0.01$ &  & $23.34 \pm 0.01$ & $2.75 \pm 0.02$ & $25.11 \pm 0.01$ & $4.56 \pm 0.03$ & $0.75 \pm 0.02$ \\ 
C-6-31 & 12:48:52.5 & $-$41:32:24.5 & $19.64 \pm 0.03$ & $0.88 \pm 0.05$ & $23.06 \pm 0.02$ & $2.40 \pm 0.02$ & $24.92 \pm 0.09$ & $4.13 \pm 0.02$ & $0.65 \pm 0.01$ \\ 
C-3-32 & 12:49:38.3 & $-$41:23:57.5 & $19.95 \pm 0.01$ & $0.98 \pm 0.02$ & $22.71 \pm 0.02$ & $1.27 \pm 0.01$ & $24.43 \pm 0.01$ & $1.99 \pm 0.02$ & $1.50 \pm 0.05$ \\ 
C-4-33 & 12:49:46.1 & $-$41:17:56.7 & $19.99 \pm 0.05$ & $0.94 \pm 0.07$ & $24.17 \pm 0.01$ & $3.53 \pm 0.03$ & $25.91 \pm 0.02$ & $5.68 \pm 0.07$ & $0.89 \pm 0.01$ \\ 
C-2-34 & 12:49:20.9 & $-$41:17:11.0 & $20.00 \pm 0.03$ & $0.82 \pm 0.06$ & $23.07 \pm 0.02$ & $2.05 \pm 0.02$ & $24.90 \pm 0.01$ & $3.49 \pm 0.02$ & $0.78 \pm 0.01$ \\ 
C-2-35 & 12:49:13.2 & $-$41:17:56.0 & $20.11 \pm 0.04$ & $0.92 \pm 0.05$ & $24.11 \pm 0.03$ & $2.92 \pm 0.06$ & $25.84 \pm 0.02$ & $4.72 \pm 0.06$ & $0.78 \pm 0.02$ \\ 
C-1-36 & 12:48:37.9 & $-$41:19:48.7 & $20.15 \pm 0.03$ & $0.75 \pm 0.06$ & $23.30 \pm 0.02$ & $2.00 \pm 0.02$ & $25.11 \pm 0.02$ & $3.37 \pm 0.03$ & $0.82 \pm 0.02$ \\ 
C-2-37 & 12:49:15.1 & $-$41:17:10.9 & $20.15 \pm 0.01$ & $0.86 \pm 0.02$ & $23.35 \pm 0.04$ & $2.09 \pm 0.04$ & $24.15 \pm 0.01$ & $2.27 \pm 0.01$ & $1.13 \pm 0.01$ \\ 
C-1-38 & 12:48:40.8 & $-$41:19:48.7 & $20.21 \pm 0.04$ & $0.86 \pm 0.05$ & $24.05 \pm 0.01$ & $2.70 \pm 0.03$ & $25.72 \pm 0.03$ & $4.18 \pm 0.06$ & $0.84 \pm 0.03$ \\ 
C-4-39 & 12:49:38.9 & $-$41:16:39.5 & $20.24 \pm 0.03$ & $0.83 \pm 0.05$ & $22.81 \pm 0.01$ & $1.82 \pm 0.01$ & $24.55 \pm 0.01$ & $2.97 \pm 0.01$ & $0.85 \pm 0.01$ \\ 
C-3-40 & 12:50:02.6 & $-$41:19:48.9 & $20.59 \pm 0.01$ & $0.94 \pm 0.02$ & $22.79 \pm 0.01$ & $1.48 \pm 0.01$ & $24.50 \pm 0.01$ & $2.39 \pm 0.01$ & $0.82 \pm 0.01$ \\ 
C-5-41 & 12:48:47.4 & $-$41:23:19.8 & $20.60 \pm 0.03$ & $0.84 \pm 0.04$ & $24.29 \pm 0.03$ & $2.44 \pm 0.05$ & $25.95 \pm 0.04$ & $3.78 \pm 0.07$ & $0.78 \pm 0.03$ \\ 
C-1-42 & 12:48:45.1 & $-$41:21:06.5 & $20.62 \pm 0.04$ & $0.84 \pm 0.05$ & $23.89 \pm 0.05$ & $2.03 \pm 0.05$ & $25.92 \pm 0.02$ & $3.79 \pm 0.05$ & $0.76 \pm 0.02$ \\ 
C-3-43 & 12:50:00.9 & $-$41:19:07.9 & $20.65 \pm 0.05$ & $0.87 \pm 0.09$ & $24.07 \pm 0.03$ & $2.27 \pm 0.04$ & $25.83 \pm 0.02$ & $3.75 \pm 0.04$ & $0.66 \pm 0.02$ \\ 
C-7-44 & 12:49:02.6 & $-$41:37:05.7 & $20.72 \pm 0.05$ & $0.66 \pm 0.07$ & $23.86 \pm 0.05$ & $2.12 \pm 0.05$ & $25.67 \pm 0.02$ & $3.57 \pm 0.04$ & $0.54 \pm 0.01$ \\ 
C-5-45 & 12:48:40.6 & $-$41:24:19.3 & $20.76 \pm 0.02$ & $0.86 \pm 0.03$ & $23.03 \pm 0.02$ & $1.64 \pm 0.02$ & $24.75 \pm 0.01$ & $2.69 \pm 0.01$ & $0.61 \pm 0.01$ \\ 
C-1-46 & 12:49:00.3 & $-$41:18:48.0 & $20.78 \pm 0.04$ & $0.88 \pm 0.06$ & $23.71 \pm 0.01$ & $1.80 \pm 0.02$ & $25.56 \pm 0.05$ & $3.04 \pm 0.07$ & $1.08 \pm 0.07$ \\ 
C-1-47$^a$ & 12:48:27.2 & $-$41:17:35.1 & $20.79 \pm 0.05$ &  & $24.06 \pm 0.03$ & $2.46 \pm 0.04$ & $25.84 \pm 0.03$ & $4.07 \pm 0.06$ & $0.93 \pm 0.03$ \\ 
C-4-48 & 12:50:03.8 & $-$41:11:36.0 & $20.82 \pm 0.03$ & $0.74 \pm 0.04$ & $22.98 \pm 0.01$ & $1.27 \pm 0.01$ & $24.57 \pm 0.01$ & $1.96 \pm 0.01$ & $0.69 \pm 0.02$ \\ 
C-3-49 & 12:49:35.9 & $-$41:21:07.5 & $20.87 \pm 0.06$ & $1.03 \pm 0.07$ & $24.40 \pm 0.02$ & $3.01 \pm 0.04$ & $26.09 \pm 0.04$ & $4.71 \pm 0.10$ & $0.90 \pm 0.03$ \\ 
C-3-50 & 12:49:47.9 & $-$41:22:04.0 & $20.90 \pm 0.05$ & $0.82 \pm 0.07$ & $24.11 \pm 0.02$ & $2.07 \pm 0.03$ & $25.80 \pm 0.02$ & $3.30 \pm 0.02$ & $0.63 \pm 0.01$ \\ 
C-2-51 & 12:49:32.0 & $-$41:17:18.2 & $20.97 \pm 0.03$ & $0.79 \pm 0.04$ & $23.24 \pm 0.03$ & $1.69 \pm 0.03$ & $25.01 \pm 0.02$ & $2.81 \pm 0.03$ & $0.74 \pm 0.02$ \\ 
C-5-52 & 12:48:52.3 & $-$41:27:13.7 & $21.04 \pm 0.09$ & $1.03 \pm 0.13$ & $24.13 \pm 0.07$ & $2.17 \pm 0.06$ & $26.21 \pm 0.05$ & $4.07 \pm 0.09$ & $0.65 \pm 0.08$ \\ 
C-5-53 & 12:48:47.4 & $-$41:26:05.1 & $21.13 \pm 0.04$ & $0.87 \pm 0.06$ & $23.83 \pm 0.06$ & $1.59 \pm 0.04$ & $25.67 \pm 0.02$ & $2.73 \pm 0.03$ & $0.64 \pm 0.03$ \\ 
C-4-54 & 12:50:01.9 & $-$41:12:46.9 & $21.16 \pm 0.08$ & $0.91 \pm 0.12$ & $24.49 \pm 0.04$ & $2.30 \pm 0.05$ & $26.25 \pm 0.04$ & $3.77 \pm 0.07$ & $0.84 \pm 0.03$ \\ 
C-6-55 & 12:49:01.4 & $-$41:30:17.6 & $21.19 \pm 0.01$ & $0.79 \pm 0.02$ & $23.18 \pm 0.03$ & $1.18 \pm 0.02$ & $24.85 \pm 0.02$ & $1.88 \pm 0.02$ & $0.62 \pm 0.03$ \\ 
C-7-56 & 12:48:32.6 & $-$41:35:41.9 & $21.27 \pm 0.22$ & $0.77 \pm 0.31$ & $25.79 \pm 0.03$ & $4.64 \pm 0.14$ & $27.12 \pm 0.03$ & $6.04 \pm 0.10$ & $0.46 \pm 0.02$ \\ 
C-4-57 & 12:50:05.1 & $-$41:12:38.8 & $21.30 \pm 0.17$ & $0.69 \pm 0.21$ & $24.67 \pm 0.01$ & $2.32 \pm 0.03$ & $26.61 \pm 0.06$ & $4.12 \pm 0.13$ & $1.14 \pm 0.06$ \\ 
C-3-58 & 12:49:40.8 & $-$41:21:06.5 & $21.34 \pm 0.08$ & $0.84 \pm 0.10$ & $24.75 \pm 0.03$ & $2.21 \pm 0.06$ & $26.35 \pm 0.04$ & $3.34 \pm 0.08$ & $0.76 \pm 0.04$ \\ 
C-3-59 & 12:49:57.9 & $-$41:19:22.2 & $21.38 \pm 0.03$ & $0.76 \pm 0.04$ & $23.18 \pm 0.01$ & $1.24 \pm 0.01$ & $24.87 \pm 0.01$ & $1.98 \pm 0.01$ & $0.75 \pm 0.02$ \\ 
C-1-60 & 12:48:36.5 & $-$41:18:37.1 & $21.48 \pm 0.04$ & $0.80 \pm 0.10$ & $24.58 \pm 0.04$ & $2.78 \pm 0.08$ & $26.25 \pm 0.05$ & $4.34 \pm 0.10$ & $0.78 \pm 0.05$ \\ 
C-2-61 & 12:49:08.2 & $-$41:21:13.8 & $21.50 \pm 0.04$ & $0.91 \pm 0.07$ & $25.09 \pm 0.03$ & $5.93 \pm 0.33$ & $26.35 \pm 0.10$ & $6.31 \pm 0.40$ & $0.69 \pm 0.06$ \\ 
C-6-62 & 12:48:59.3 & $-$41:30:19.3 & $21.52 \pm 0.02$ & $0.82 \pm 0.04$ & $23.69 \pm 0.04$ & $1.25 \pm 0.03$ & $25.35 \pm 0.04$ & $1.96 \pm 0.04$ & $0.72 \pm 0.04$ \\ 
C-2-63 & 12:49:07.1 & $-$41:19:38.4 & $21.62 \pm 0.04$ & $0.98 \pm 0.06$ & $23.77 \pm 0.15$ & $1.25 \pm 0.07$ & $25.95 \pm 0.05$ & $2.49 \pm 0.07$ & $0.79 \pm 0.04$ \\ 
C-4-64 & 12:49:58.0 & $-$41:15:11.6 & $21.75 \pm 0.10$ & $0.95 \pm 0.15$ & $25.38 \pm 0.03$ & $2.50 \pm 0.06$ & $27.06 \pm 0.08$ & $3.94 \pm 0.16$ & $0.81 \pm 0.06$ \\ 
C-4-65 & 12:50:06.5 & $-$41:13:39.7 & $21.83 \pm 0.04$ & $0.90 \pm 0.09$ & $24.11 \pm 0.02$ & $1.68 \pm 0.02$ & $25.95 \pm 0.05$ & $2.84 \pm 0.07$ & $0.99 \pm 0.07$ \\ 
C-1-66 & 12:48:34.1 & $-$41:21:20.8 & $21.92 \pm 0.12$ & $0.84 \pm 0.16$ & $25.07 \pm 0.15$ & $2.07 \pm 0.16$ & $26.81 \pm 0.10$ & $3.37 \pm 0.18$ & $0.59 \pm 0.08$ \\ 
C-3-67 & 12:49:54.7 & $-$41:19:00.7 & $21.94 \pm 0.05$ & $1.03 \pm 0.07$ & $24.12 \pm 0.10$ & $1.42 \pm 0.07$ & $26.12 \pm 0.05$ & $2.59 \pm 0.06$ & $0.75 \pm 0.05$ \\ 
C-5-68 & 12:48:47.3 & $-$41:22:05.1 & $21.98 \pm 0.03$ & $0.91 \pm 0.05$ & $24.23 \pm 0.02$ & $2.41 \pm 0.05$ & $25.83 \pm 0.05$ & $3.56 \pm 0.10$ & $0.85 \pm 0.04$ \\ 
C-3-69 & 12:49:42.0 & $-$41:21:33.0 & $22.18 \pm 0.11$ & $0.86 \pm 0.15$ & $25.02 \pm 0.07$ & $1.77 \pm 0.08$ & $26.70 \pm 0.09$ & $2.79 \pm 0.15$ & $0.78 \pm 0.09$ \\ 
C-6-70 & 12:48:45.5 & $-$41:30:18.8 & $22.24 \pm 0.04$ & $0.65 \pm 0.04$ & $23.86 \pm 0.02$ & $0.97 \pm 0.02$ & $25.64 \pm 0.09$ & $1.58 \pm 0.09$ & $0.97 \pm 0.06$ \\ 
C-4-71 & 12:49:39.4 & $-$41:12:21.1 & $22.28 \pm 0.09$ & $0.88 \pm 0.13$ & $25.01 \pm 0.05$ & $1.71 \pm 0.06$ & $26.85 \pm 0.11$ & $2.89 \pm 0.15$ & $0.94 \pm 0.12$ \\ 
C-3-72 & 12:50:00.9 & $-$41:21:32.5 & $22.33 \pm 0.05$ & $0.93 \pm 0.08$ & $24.28 \pm 0.05$ & $1.24 \pm 0.04$ & $25.99 \pm 0.06$ & $2.00 \pm 0.07$ & $0.79 \pm 0.08$ \\ 
C-3-73 & 12:49:59.6 & $-$41:20:16.1 & $22.37 \pm 0.07$ & $0.67 \pm 0.15$ & $24.38 \pm 0.09$ & $1.35 \pm 0.01$ & $26.05 \pm 0.08$ & $2.14 \pm 0.08$ & $0.60 \pm 0.20$ \\ 
C-1-74 & 12:48:39.4 & $-$41:17:00.0 & $22.41 \pm 0.06$ & $0.67 \pm 0.11$ & $24.78 \pm 0.06$ & $1.58 \pm 0.06$ & $26.66 \pm 0.11$ & $2.73 \pm 0.16$ & $1.05 \pm 0.11$ \\ 
C-4-75 & 12:49:42.8 & $-$41:13:12.6 & $22.42 \pm 0.06$ & $0.74 \pm 0.07$ & $24.23 \pm 0.05$ & $1.12 \pm 0.03$ & $26.02 \pm 0.09$ & $1.85 \pm 0.07$ & $0.98 \pm 0.21$ \\ 
C-1-76 & 12:48:42.0 & $-$41:18:02.6 & $22.57 \pm 0.06$ & $0.58 \pm 0.08$ & $24.09 \pm 0.08$ & $0.96 \pm 0.04$ & $25.90 \pm 0.08$ & $1.59 \pm 0.07$ & $1.00 \pm 0.19$ \\ 
C-3-77 & 12:49:41.5 & $-$41:18:34.6 & $22.71 \pm 0.13$ & $0.81 \pm 0.13$ & $24.93 \pm 0.16$ & $1.48 \pm 0.12$ & $26.57 \pm 0.07$ & $2.37 \pm 0.07$ & $0.76 \pm 0.09$ \\ 
C-1-78 & 12:48:37.6 & $-$41:17:16.7 & $22.77 \pm 0.12$ & $1.03 \pm 0.18$ & $25.25 \pm 0.09$ & $1.56 \pm 0.10$ & $27.02 \pm 0.13$ & $2.58 \pm 0.18$ & $0.87 \pm 0.16$ \\ 
C-2-79 & 12:49:05.6 & $-$41:16:43.4 & $22.80 \pm 0.06$ & $0.93 \pm 0.07$ & $25.19 \pm 0.14$ & $1.55 \pm 0.27$ & $27.10 \pm 0.59$ & $2.73 \pm 1.68$ & $1.06 \pm 0.55$ \\ 
C-4-80 & 12:49:45.9 & $-$41:14:52.7 & $22.80 \pm 0.16$ & $0.79 \pm 0.26$ & $25.51 \pm 0.14$ & $1.80 \pm 0.14$ & $27.76 \pm 0.42$ & $3.99 \pm 1.23$ & $0.91 \pm 0.43$ \\ 
C-5-81 & 12:48:49.6 & $-$41:23:01.5 & $22.86 \pm 0.05$ & $0.70 \pm 0.08$ & $24.46 \pm 0.07$ & $1.16 \pm 0.05$ & $26.31 \pm 0.16$ & $1.97 \pm 0.17$ & $1.00 \pm 0.25$ \\ 
C-2-82 & 12:49:06.9 & $-$41:18:48.5 & $22.90 \pm 0.10$ & $0.84 \pm 0.14$ & $25.61 \pm 0.14$ & $1.54 \pm 0.22$ & $27.03 \pm 0.12$ & $1.98 \pm 0.20$ & $0.80 \pm 0.02$ \\ 
C-2-83 & 12:49:25.4 & $-$41:18:21.9 & $22.98 \pm 0.12$ & $0.88 \pm 0.15$ & $25.69 \pm 0.08$ & $2.53 \pm 0.18$ & $27.39 \pm 0.43$ & $3.96 \pm 1.30$ & $0.92 \pm 0.40$ \\ 
C-1-84 & 12:48:31.1 & $-$41:15:34.4 & $23.01 \pm 0.07$ & $0.75 \pm 0.08$ & $24.97 \pm 0.11$ & $1.09 \pm 0.07$ & $27.02 \pm 0.15$ & $2.03 \pm 0.20$ & $1.39 \pm 0.11$ \\ 
C-2-85 & 12:49:24.6 & $-$41:15:14.5 & $23.13 \pm 0.14$ & $0.64 \pm 0.19$ & $24.91 \pm 0.18$ & $1.16 \pm 0.15$ & $26.75 \pm 0.22$ & $2.00 \pm 0.35$ & $0.93 \pm 0.16$ \\ 
C-1-86 & 12:48:40.0 & $-$41:19:17.4 & $23.17 \pm 0.06$ & $0.53 \pm 0.09$ & $24.63 \pm 0.12$ & $1.12 \pm 0.10$ & $26.44 \pm 0.17$ & $1.88 \pm 0.20$ & $0.79 \pm 0.21$ \\ 
C-7-87 & 12:48:42.4 & $-$41:36:52.1 & $23.21 \pm 0.19$ & $0.41 \pm 0.27$ & $25.41 \pm 0.09$ & $1.38 \pm 0.14$ & $27.29 \pm 0.26$ & $2.44 \pm 0.47$ & $1.02 \pm 0.25$ \\ 
C-5-88 & 12:48:28.3 & $-$41:21:24.5 & $23.37 \pm 0.10$ & $0.82 \pm 0.11$ & $25.01 \pm 0.16$ & $1.03 \pm 0.11$ & $26.90 \pm 0.45$ & $1.79 \pm 0.70$ & $1.00 \pm 0.37$ \\ 
C-7-89 & 12:48:45.2 & $-$41:36:57.3 & $23.40 \pm 0.04$ & $0.90 \pm 0.04$ & $24.47 \pm 0.16$ & $0.78 \pm 0.08$ & $26.25 \pm 0.25$ & $1.28 \pm 0.22$ & $0.99 \pm 0.25$ \\ 
C-6-90 & 12:48:59.5 & $-$41:32:02.1 & $23.44 \pm 0.14$ & $0.65 \pm 0.19$ & $25.23 \pm 0.05$ & $1.12 \pm 0.05$ & $27.00 \pm 0.12$ & $1.84 \pm 0.13$ & $0.93 \pm 0.09$ \\ 
C-5-91 & 12:49:02.2 & $-$41:22:47.5 & $23.57 \pm 0.04$ & $0.80 \pm 0.05$ & $24.75 \pm 0.11$ & $1.01 \pm 0.10$ & $27.04 \pm 0.17$ & $2.41 \pm 0.27$ & $1.19 \pm 0.13$ \\ 
C-2-92 & 12:49:08.2 & $-$41:18:08.3 & $23.88 \pm 0.12$ & $0.86 \pm 0.15$ & $25.25 \pm 0.15$ & $1.30 \pm 0.17$ & $26.89 \pm 0.20$ & $2.01 \pm 0.30$ & $0.76 \pm 0.14$ \\

\end{longtable}
\noindent
$^a$ galaxies for which no colour could be measured; $^b$ galaxies showing a two component surface brightness profile, not well fitted by a single S\'ersic law \\
$^c$ NGC 4696; $^d$ NGC 4709; $^e$ NGC 4706;
}
}

\end{appendix}

\end{document}